\documentclass[aps,pra,groupedaddress,preprint]{revtex4}
\usepackage{amsfonts, amssymb, amsmath, graphicx, comment, bm, slashed, caption, color} 
\usepackage{ulem}

\newcommand \beq{\begin{equation}}
\newcommand \eeq{\end{equation}}

\newcommand{\Slash}[1]{{\ooalign{\hfil/\hfil\crcr$#1$}}}
\newcommand{\tr}{{\rm tr}}
\newcommand{\Tr}{{\rm Tr}}
\newcommand{\Ln}{{\rm Ln}}

\newcommand{\vp}{ {\bf p}}
\newcommand{\vq}{ {\bf q}}
\newcommand{\vk}{ {\bf k}}
\newcommand{\vl}{ {\bf l}}

\newcommand{\para}{\parallel}

\newcommand{\calG}{\mathcal{G}}
\newcommand{\calI}{\mathcal{I}}

\newcommand{\calM}{\mathcal{M}}

\newcommand{\calV}{\mathcal{V}}

\newcommand{\rmd}{\mathrm{d}}
\newcommand{\rmi}{\mathrm{i}}
\newcommand{\rme}{\mathrm{e}}

\newcommand{\rmpp}{ {\rm p } }
\newcommand{\rma}{ {\rm a} }
\newcommand{\tp}{ \tilde{p} }

\newcommand{\tl}{ \tilde{l} }

\begin{document}
\title{Zero point energy of composite particles: 
The medium effects}
\author{Toru Kojo}
\affiliation{Key Laboratory of Quark and Lepton Physics (MOE) and Institute of Particle Physics, Central China Normal University, Wuhan 430079, China}  
\date{\today}

\begin{abstract}
We analyze the zero point energy of composite particles (or resonances) which are dynamically created from relativistic fermions.  We compare the zero point energies in medium to the vacuum one, taking into account the medium modification of the constituent particles.
Treating composite particles as if quasi-particles, their zero point energies contain the quadratic and logarithmic UV divergences even after the vacuum subtraction. 
The coefficients of these divergences come from the difference between the vacuum and in-medium fermion propagators. 
We argue that such apparent divergences can be cancelled by consistently using fermion propagators to compute the quasi-particle contributions as well as their interplay, provided that the self-energies of the constituents at large momenta approach to the vacuum ones sufficiently fast. 
In the case of quantum chromodynamics, mesons and baryons, which may be induced or destroyed by medium effects, yield the in-medium divergences in the zero point energies, but the divergences are assembled to cancel with those from the quark zero-point energy. 
This is particularly important for unified descriptions of hadronic and quark matter which may be smoothly connected by the quark-hadron continuity

\end{abstract}

\maketitle

\section{Introduction}
\label{sec:intro}

Strongly correlated systems of fermions typically develop composite objects or collective modes \cite{bohr,Nambu:1961tp,Higgs}. The examples include fermion pairs in superfluidity \cite{AGD}, composite fermions in quantum hall systems \cite{composite_fermon}, hadrons in quantum chromodynamics (QCD), and many others. Those states often play dominant roles in transport phenomena or in thermal equations of state, especially when fermions as constituents have the excitation energies larger than the composites. 

In principle the composites can also contribute to the zero point energy, i.e., the vacuum energy of a system or equations of state at zero temperature through their quantum fluctuations. The consideration for these contributions may be potentially important for, e.g., QCD equations of state at high baryon density where the relevant degrees of freedom change from hadrons to quarks \cite{Schafer:1998ef,Baym:2019iky,Baym:2017whm,Masuda:2012kf,Kojo:2014rca,Fukushima:2015bda,McLerran:2018hbz,Jeong:2019lhv,Ma:2019ery}. This paper will address problems of the in-medium zero point energy of composite particles. The UV divergences appear if the interplay between the composites and their constituents is not properly taken into account. 

The UV divergences to be discussed are associated with the summation of states for composite objects. Composite particles have the total momenta which must be integrated. One might think that  the composite particle dissociates at some momentum which provides a cutoff on the integral. But in general there are also states whose energy spectra survive to high energy and behave as $E_{ {\rm comp} } (\vk) \sim \sqrt{ m_{ {\rm comp} } ^2 + v^2 \vk^2 \,}$  ($m_{ {\rm comp} } $: the energy in the rest frame; $\vk$: spatial momentum; $v$: velocity approaching to the light velocity in the UV limit). 
The number of their states grows as $\sim |\vk|^3$; in this case the zero point energy is $\sim \int \rmd^3 \vk \, E_{ {\rm comp} } (\vk) = a \Lambda_{ {\rm UV} }^4 + b m_{ {\rm comp} } ^2 \Lambda_{ {\rm UV} }^2 + c m_{ {\rm comp} } ^4 \ln \Lambda_{ {\rm UV} } + \cdots$, where $\Lambda_{ {\rm UV}}$ is some UV scale much bigger than the natural scale of the theory. The first term is saturated by the physics at high energy and hence is universal, so can be eliminated by the vacuum subtraction. This is not the case for the second and the third terms which arise from the coupling between universal hard part and soft dynamics; the soft part can be easily affected by the environment and is not universal. 
These arguments clearly require closer inspections based on more microscopic degrees of freedom.

Another problem is that, going from zero to high fermion density, composite states may not keep the one-to-one correspondence to their vacuum counterparts, 
e.g., some composite states in vacuum might disappear in medium. This mismatch in the degrees of freedom likely leaves the mismatch in the UV contributions which in turn appear as the UV divergences in equations of state. A more general question is how excited states, including the resonances and continuum, should be taken into account. To discuss these issues, some integral representations should be used to include all possible states in a given channel \cite{Dashen:1969ep,Blaschke:2013zaa,Bastian:2018wfl,Kojo:2017gxc,Kojo:2017opq}. 

In this paper we analyze the structure of UV divergences which come from the composite particles and their constituents. We argue that, to handle the UV divergences associated with the composite particles,  it is most essential to take into account the interplay between the composites and their constituents. In short, they transfer the UV divergences one another when the fermion bases (or propagators) are deformed in a non-perturbative way. As a consequence, graph by graph cancellations by the vacuum subtraction do not take place; only the sum cancels. 
To keep track the impact of changes in fermion bases, we apply the two-particle irreducible (2PI)-formalism.

In order to make the zero point energy UV finite, we demand: (i) all {\it vacuum} $n$-point functions are made finite in some way; (ii) the fermion self-energies approach the universal limit sufficiently fast at high energy.
With these conditions the quadratic divergences in the zero-point energy cancel, as we can see from the general structure of the 2PI functional.
Meanwhile the removal of the logarithmic divergences depend on models and requires more sophisticated discussions, but after all the origin of the problems can be traced back to the in-medium self-energies.

The discussions were originally motivated for the applications to QCD equations of state at finite baryon density \cite{Baym:2017whm,Masuda:2012kf,Kojo:2014rca,Fukushima:2015bda,McLerran:2018hbz,Jeong:2019lhv,Ma:2019ery,Freedman:1976xs,Kurkela:2009gj,Xia:2019xax,Han:2019bub}. But we consider the problems in more general context, aiming at not only renormalizable theories but also non-renormalizable ones. 
The latter is often used for practical calculations of equations of state, but usually requires introduction of a UV cutoff which always leaves questions concerning with impacts of the cutoff effects or artifacts. 
In particular whether equations of state depend on the model cutoff quadratically or logarithmically makes important difference. 
It is important to identify the origins; the strong cutoff dependences may be intrinsic to the structure of models whose origin can be traced back to the microphysics, but some may be due to inconsistent approximations and are artifacts to be eliminated. In this paper we try to structure our arguments in such a way that we will be able to disentangle the intrinsic and artificial cutoff dependence.

To make our arguments concrete, we will often refer to the gaussian pair-fluctuation theory in which the fermion pair fluctuations are added on top of mean field results. For non-relativistic fermions, this theory has been successful descriptions in the context of the BEC-BCS crossover \cite{NSR,Diener,Ohashi:2002zz}. On the other hand, these theories have inconsistency in the treatments of the single-particle and of composite particles, and, in the case of relativistic fermions \cite{Abuki:2006dv,Sun:2007fc}, it leads to the UV divergences which must be cutoff by hand. It is rather easy to see the existence of the inconsistency. On the other hand, it is less obvious that such inconsistency can be the source of the UV divergences.

This paper is organized as follows. 
Sec.\ref{sec:pair} is devoted to further illustration of the UV problem. We consider fermion pairs as the simplest example of composites and discuss their contributions to the equations of state. 
In Sec.\ref{sec:single} we discuss the ``single particle" contribution with the self-energy and its UV divergent contribution to the equation of state. 
In Sec.\ref{sec:2PI} we briefly overview the structure of the 2PI-functional and present the general strategy to handle the UV divergences. We isolate the medium contributions induced by the medium modifications of fermion propagators.
In Sec.\ref{sec:same_bases} we first discuss the medium contributions without modifications of the vacuum fermion bases.
In Sec.\ref{sec:change_bases} we examine the impacts of the change of fermion bases and discuss how to handle the UV divergences associated with it. The interplay between single particle's and composite's contributions is discussed within the 2PI-formalism. 
Sec \ref{sec:summary} is devoted to summary.

 We use the notation $\int_x = \int \rmd^4 x$, $\int_p = \int \frac{\rmd^4 p}{\, (2\pi)^4 \,}$, $\int_\vp = \int \frac{\rmd^3 \vp}{\, (2\pi)^3 \,}$, and $\Tr_N [\cdots]= \left( \Pi_{n=1}^N \int_{p_n} \right) \tr_{D}[\cdots]$ where $\tr_{D}$ are trace over the Dirac indices. The momenta $\tp_\mu = (p_0-\rmi \mu, p_j)$ will be also used.

\section{The pair fluctuations}
\label{sec:pair}

\begin{figure*}[!t]
\begin{center}
\vspace{-1.5cm}
\hspace{-.3cm}
\includegraphics[width = 0.5\textwidth]{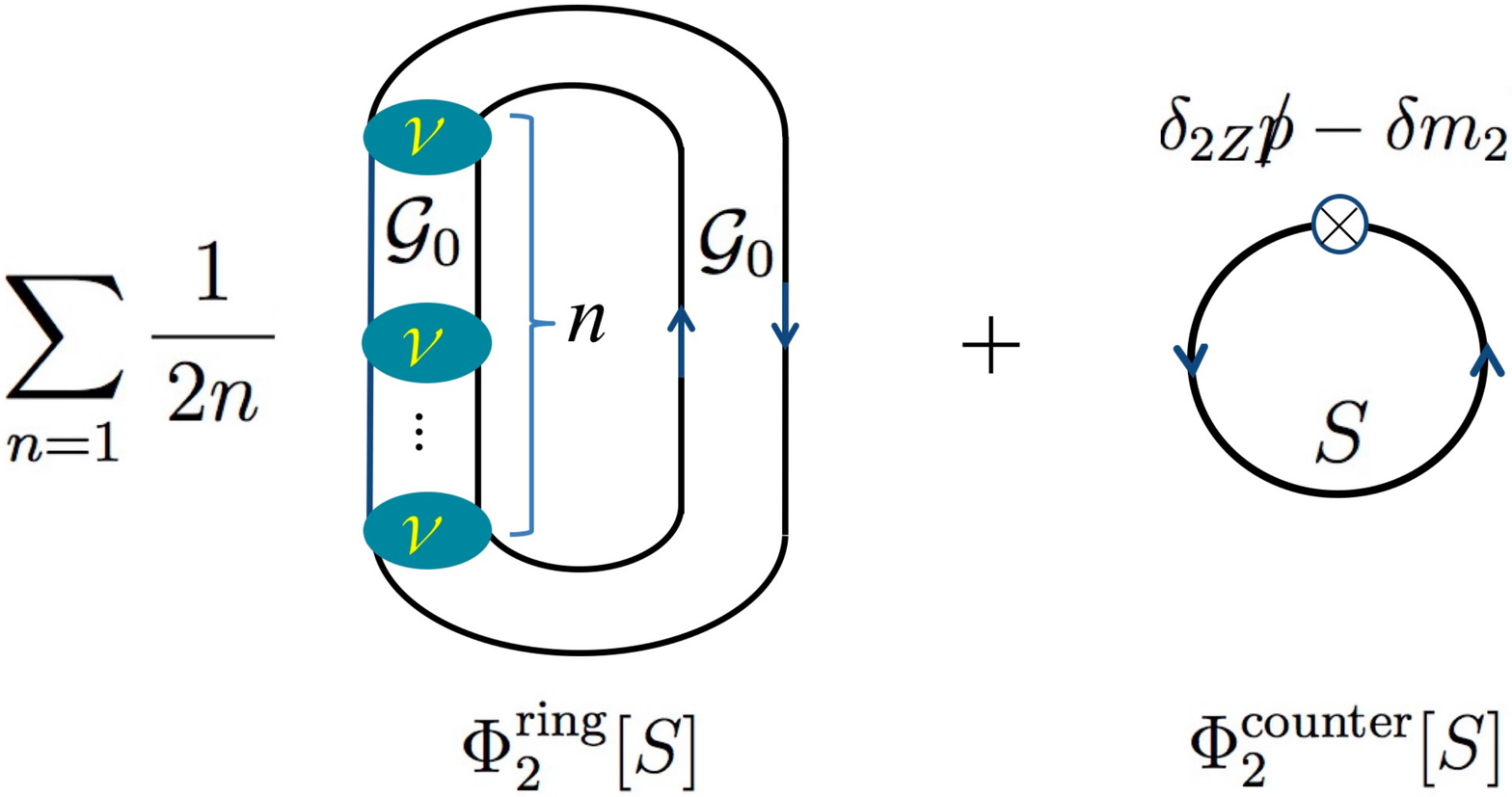}
\vspace{-0.cm}
\end{center}
\vspace{-1.5cm}
\caption{ 
\footnotesize{
The pair fluctuation diagrams that generate quadratic divergences. The counter terms, whose values are fixed in the vacuum calculations, are also shown.}
\vspace{-0.0cm} }
\label{fig:pair_graph}
\end{figure*}

As the simplest example, we consider the fermion pair-fluctuation contributions to the equations of state. The resummation of 2-body graphs generates bound states, resonance poles, and continuum. The popular approach is the gaussian pair-fluctuation theory in which the contributions from the 2-particle correlated contributions are simply added to the mean field equations of state. In this section we examine how the UV divergence appears in equations of state in theories of relativistic fermions.

We start with a free streaming 2-quasiparticle propagator,
\beq
\calG^0_{\alpha\alpha' ; \beta \beta'} ( p; k ) = S_{\alpha \alpha'} ( p_+) S_{\beta \beta'} ( p_- ) \,, ~~~~~~~~ p_\pm = p \pm \frac{\, k \,}{2} \,,
\eeq
where $k$ is the total momentum of the two particle which is conserved, while $p$ is the relative momentum. The self-energy corrections are included into the definition of $S$. We write the 2-body interaction as $\calV$, then the resummed 2-fermion propagator is given in a symbolic form as
\beq
\calG = \calG_{0} +  \calG_{0} \calV \calG_{0} + \cdots = \frac{ \calG_0 }{\, 1 - \calV \calG_0 \,} \,,
\eeq
where the summation over internal momenta and other quantum numbers is implicit. The poles are found by searching for the values of $k_\mu = (E(\vk),\vk)$  such that $1-\calV \calG_0=0$.

We assume that $\calG$ and $\calV$ {\it in vacuum} are made finite by some renormalization procedures or by introduction of physical cutoffs. For theories like QCD or QED, the 4-fermion functions are UV finite as far as we use the renormalized fermion propagators and vertices. For theories with contact interactions the 4-fermion functions are UV divergent and must be made finite by physically motivated cutoffs \cite{Hatsuda:1994pi}. In this paper we call such cutoffs intrinsic to models 
``physical cutoff", and distinguish it from $\Lambda_{ {\rm UV} }$, a regulator which must be taken to infinity at the end of calculations. After $n$-point functions are made finite, $\Lambda_{ {\rm UV} }$ shows up only when we close all the lines of $n$-point functions to compute the zero-point energy.

In the expression for equations of state, this 2-fermion propagator appears inside of the logarithm; drawing the corresponding closed diagrams  (Fig.\ref{fig:pair_graph}, left), we must attach the symmetry factor $1/(2n)$ and sum them up,
\beq
\Phi^{ {\rm ring} }_{ {\rm 2} } [S] = \sum_{n=1} \frac{1}{\, 2n \,} \Tr_2\left[  \left( \calV \calG_0 \right)^n \right]
 = - \frac{1}{\, 2 \,} \Tr_2 \left[ \Ln \left( 1- \calV \calG_0 \right) \right] 
 = \frac{1}{\, 2 \,} \Tr_2 \Ln  \left( \calG/\calG_0 \right)^{-1} \,,
 \label{eq:gauss}
\eeq
where we wrote the energy density as a functional of $S$ for the later purpose. The expression includes bound states, resonances, and the continuum in the 2-body channel. 

The form of $\Ln ( \calG/\calG_0 )$ reflects the fact that uncorrelated product of fermion lines in $\calG$ must be subtracted to avoid the double counting (linked-cluster theorem). It should be also clear that  the term $\Ln ( \calG/\calG_0 )$ approaches zero at very large energy; the kinetic energy is much larger than the potential energy and hence $\calG \rightarrow \calG_0$. Meanwhile, even though this cancels the leading UV contributions, we expect $\Ln ( \calG/\calG_0 ) \sim 1/k^2$ that may produce the quadratic and logarithmic divergences after integrating the momentum $k$.

We have not yet introduced a counter term or a vacuum subtraction procedure for the zero point function, so we set our reference energy density to the vacuum one, $\Phi^{ {\rm ring} }_2 [S_{ {\rm vac} }] $, evaluated for some propagator $S_{ {\rm vac} }$ at $\mu=0$. Then the renormalized energy density is
\beq
 \Phi_{ {\rm 2R} }^{{\rm ring} } [S] =  \Phi^{ {\rm ring} }_{ {\rm 2} } [S] - \Phi^{ {\rm ring} }_{2} [S_{ {\rm vac} }] \,.
\eeq
However, this vacuum subtraction in general is not sufficient to cancel the above-mentioned quadratic and logarithmic divergences. The coefficients of $\sim \Lambda_{ {\rm UV} }^2$ and $ \sim \ln \Lambda_{ {\rm UV} }$ terms are sensitive to the physics in the IR, and can differ for $S$ and $S_{ {\rm vac} }$. 

\begin{figure*}[!t]
\begin{center}
\vspace{-1.5cm}
\hspace{-.3cm}
\includegraphics[width = 0.6\textwidth]{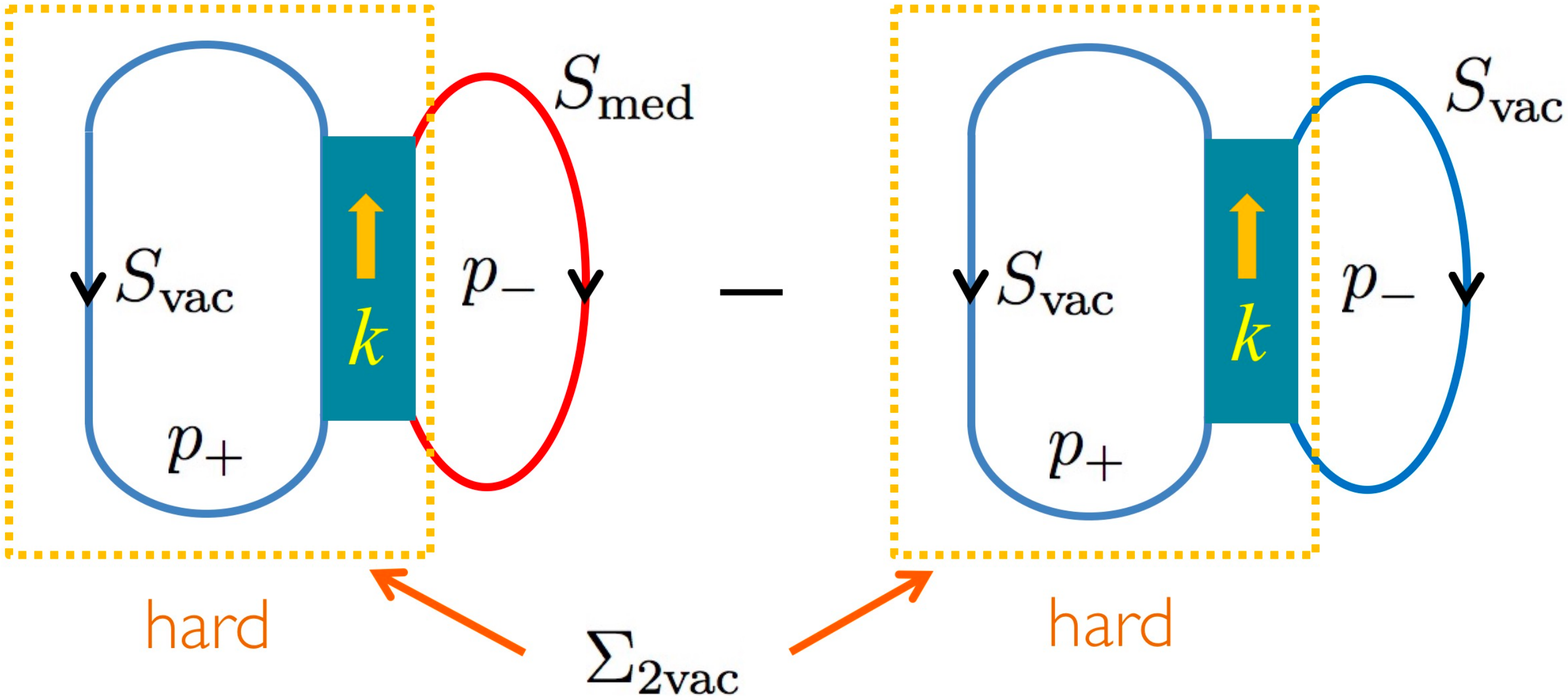}
\vspace{-0.cm}
\end{center}
\vspace{-1.5cm}
\caption{ 
\footnotesize{The difference between the medium and vacuum self-energy from the 2-particle correlations. Very large momenta, $p_+,k\rightarrow \infty$, flows into the graph, except a fermion line with momentum $p_-$ which is finite. (The counter terms to renormalize the vacuum self-energy part is not explicitly shown.) }
\vspace{-0.0cm} }
\label{fig:dif_2particle}
\end{figure*}

To see how these UV divergences arise, let us consider for the moment the diagram such that the graphs except the propagator carrying momentum $p_-$ is set to $S_{ {\rm vac} }$, see Fig.\ref{fig:dif_2particle}. We consider the case where large momentum $k$ flows into the two-body scattering kernel, as we are interested in the composites at large total momenta. This contribution has the structure,
\beq
\sim \int_{p_-} \Sigma_{2{\rm vac} } (p_-) \big( S_{ {\rm med} }  (p_-) - S_{ {\rm vac} }  (p_-) \big) 
\sim - \int_{p_-} \Sigma_{2{\rm vac} } S_{ {\rm vac} }\big( \Sigma_{ {\rm med} }  - \Sigma_{ {\rm vac} } \big) S_{ {\rm vac} }
\,,  
\eeq
where we expand $S_{ {\rm med} } \simeq S_{ {\rm vac} } - S_{ {\rm vac} } \big( \Sigma_{ {\rm med} }  - \Sigma_{ {\rm vac} } \big) S_{ {\rm vac} }$.
(There is also the integral over $p_+$, but it is hidden in subgraphs of the vacuum self-energy $ \Sigma_{ 2{\rm vac} } (p_-)$ multiplied to $ S_{ {\rm med} } - S_{ {\rm vac} }$.)
 At this point in treating the vacuum self-energy in the subgraph, it is necessary to mention the necessity to include counter terms \cite{Weinberg,Peskin} (Fig.\ref{fig:pair_graph}, right),
\beq
\Phi^{ {\rm counter} }_{ {\rm 2} } [S] = \Tr_1 \left[ (\delta_{2Z} \Slash{p} - \delta m_2 ) S \right] \,,
\label{eq:counter}
\eeq
which cancel the divergence in $ \Sigma_{ 2{\rm vac} } $. We note that $\delta_{2Z}$ and $\delta m_2$ are medium independent as they should, but can couple to the medium-dependent propagators $S$. These counter terms renormalize the vacuum self-energy part in the graph. Then we can focus on the medium effects coupled to the UV finite self-energy. We count the power of momenta as $\Sigma_{ 2{\rm vac} }  \sim \Sigma_{ {\rm vac} } \sim \Sigma_{ {\rm med} } \sim p$ and $ S_{ {\rm vac} }  \sim p^{-1}$. It is plausible that the medium effects decouple at high momenta, so we expect the cancellation of the leading component 
in $\Sigma_{ {\rm med} }  - \Sigma_{ {\rm vac} }$; then the rest behaves as $ \Sigma_{ {\rm med} }  - \Sigma_{ {\rm vac} } \sim \Lambda_{ {\rm med} }^2 /p$ where $\Lambda_{ {\rm med} }$ 
is some IR scale in the medium. But even after this cancellation happens, the integral over $p$ still leads to $\sim \Lambda_{ {\rm UV} }^2 \Lambda_{ {\rm med} }^2$. 

The severest UV divergence comes from the vector self-energies and is quadratic. If we ignore the modification of the vector self-energies as in typical mean field model treatments, we have less divergence but it is still $\sim \Lambda_{ {\rm med} }^4  \ln\Lambda_{ {\rm UV} }$.
Unless we assume unreasonablly stronger damping for $ \Sigma_{ {\rm med} }  - \Sigma_{ {\rm vac} }$, we cannot avoid the UV divergences.

As we will see, the divergences associated with the product of IR and UV contributions will be handled by taking into account the interplay between a composite state and the constituents in it.

\section{The single particle contribution}
\label{sec:single}

In this section we look at the single particle contribution to the equations of state. We assume that a fermion acquires the self-energy from the IR dynamics which is sensitive to the presence of the medium. After calculating the in-medium self-energy, the last step is to close the fermion line and we get the single particle contribution to equations of state. 

We begin with the case in which the fermion acquires the momentum independent mass gaps, $M_{{\rm med}}$ at finite $\mu$, and $M_{{\rm vac}}$ at $\mu=0$. Using the propagators, $S_{ {\rm med, vac} } =-(\Slash{\tp}-M_{ {\rm med, vac} } )^{-1}$, the contribution at $\mu$ is \cite{Kapusta}
\beq
-\Tr_1 \Ln S_{ {\rm med} }^{-1}  = - 2 \int_{\vp} \left[\, \theta \left( E_{ {\rm med} } (\vp) - \mu \right) + 1 \, \right]\, E_{ {\rm med} } (\vp) \,,
~~~~~~
E_{ {\rm med} } (\vp) = \sqrt{ M_{ {\rm med} }^2 + \vp^2 \,} \,,
\eeq
where the particle and anti-particle contributions are included. After subtracting the vacuum counterpart, we get
\beq
-\Tr_1 \Ln S_{ {\rm med} }^{-1} + \Tr_1 \Ln S_{ {\rm vac} }^{-1} 
= 2 \int_{\vp} \theta \big( \mu - E_{ {\rm med} } (\vp) \big) E_{ {\rm med} } (\vp) 
- 4 \int_{\vp}  \big( E_{ {\rm med} } (\vp) - E_{ {\rm vac} } (\vp) \big) \,  \,,
\eeq
where $E_{ {\rm vac} } (\vp) = \sqrt{ M_{ {\rm vac} }^2 + \vp^2 \,}$. The last set of terms vanish if $M_{ {\rm med} } = M_{ {\rm vac} }$, but otherwise it leaves the quadratic divergence, $\sim -(M^2_{ {\rm med} } - M^2_{ {\rm vac} } ) \Lambda_{ {\rm UV} }^2$. 

To improve the situation, it is tempting to consider the momentum dependence of the gap as $M \rightarrow M(\vp)$, and demand that beyond some physical damping scale $\Lambda_{ {\rm damp} }$ the mass function $M(\vp)$ approaches to some universal value independent of the IR physics. For instance one can think of
\begin{align}
M_{ {\rm med}, {\rm vac} }^2(\vp) \sim 
 \left\{
\begin{matrix}
~ \calM_{ {\rm med}, {\rm vac} }^2  ~~~~~~ &(|\vp|^2 \le \Lambda_{ {\rm damp} }^2 )\\
~ m^2_{ {\rm univ} } + c_{ {\rm med}, {\rm vac} }/\vp^2 ~~~~~~ & \,(|\vp|^2 \ge \Lambda_{ {\rm damp} }^2 )\
\end{matrix}
\right.  \,,
\end{align}
then the zero point energy is
\beq
\int_{\vp}  \big( E_{ {\rm med} } (\vp) - E_{ {\rm vac} } (\vp) \big) ~\sim~ \left( \calM_{ {\rm med}}^2 - \calM_{ {\rm vac} }^2 \right) \Lambda_{ {\rm damp} }^2 +  \left( c_{ {\rm med}}^2 - c_{ {\rm vac} }^2 \right) \ln \Lambda_{ {\rm UV} } \,.
\eeq
The first term is characterized by the damping scale. The typical model results are obtained if we entirely neglect the terms $c_{ {\rm med}, {\rm vac} }$ in the UV region, but the validity of such procedures is uncertain from the physical point of view.
In general $c_{ {\rm med}, {\rm vac} }$ exist and are non-universal, leaving the logarithmic UV divergence.

The situation gets even worse if we also take into account the modification of the residue function $Z_{ {\rm med}, {\rm vac} } (\vp)$ which appears if the vector self-energy is nonzero. The propagators are  $S_{ {\rm med, vac} } = - Z_{ {\rm med, vac} }(\Slash{\tp}-M_{ {\rm med, vac} } )^{-1}$. Assuming the damping $Z_{ {\rm med}, {\rm vac} } (\vp) \sim Z_{ {\rm univ} } + d_{ {\rm med}, {\rm vac} }/ \vp^2 $, the difference of $\Tr \Ln S$ between the medium and vacuum cases leaves the quadratic divergence $ \sim \left( d_{ {\rm med} } - d_{ {\rm vac} } \right) \Lambda_{ {\rm UV} }^2 $. 

Therefore the damping of the self-energy at high momenta alone cannot be the sufficient condition for the UV finiteness of equations of state (unless extraordinary damping takes place in the self-energy). In order to get physical equations of state, we need to assemble these UV divergent pieces with those from other origins.

\section{Equations of state in the 2PI formalism: some definitions}
\label{sec:2PI}

\subsection{The structure of 2PI functional}

We have seen that the quasi-particle contributions to equations of state from the composites and their constituents in general have the quadratic divergences. It is tempting to expect these divergences to cancel. Actually these terms do not directly cancel and we must go one step further to correctly handle the double counting problem. 

For this purpose we use the formalism of the two particle irreducible (2PI) action, developed by Luttinger-Ward \cite{Luttinger:1960ua}, Baym-Kadanoff \cite{Baym:1961zz,Baym:1962sx}, and its relativistic version by Cornwall-Jackiw-Toumbolis \cite{Cornwall:1974vz}.  
Its renormalizability was first discussed in \cite{vanHees:2001ik} and since then seminal works have followed \cite{Blaizot:2003an}. These works gave detailed account by explicitly choosing some of renormalizable theories, while we discuss the structure of the medium-induced UV divergences in more abstract fashion, so that we can emphasize the common aspects in the renormalization programs. In particular our arguments also may include the cases for non-renormalizable models. 

The 2PI action $I[S;\mu]$ is a functional of a fermion propagator $S$,
\beq
I [S;\mu] =  \Tr_1 \Ln S + \Tr_1 \left[ S \tilde{\Sigma}^{ [S;\mu] } \right] + \Phi[S] \,,~~~~~~\tilde{\Sigma}^{ [S;\mu] } = S^{-1}- ( S^{\mu}_{ {\rm tree} } )^{-1} \,,
\eeq
which includes a tree level propagator (in Euclidean space), $S^{\mu}_{ {\rm tree} } (p) = S^{\mu=0}_{ {\rm tree} } (\tp) = -\left( \Slash{\tp} - m \right)^{-1}$, with the fermion mass $m$. The $\Tr_1[S\tilde{\Sigma}]$ term plays a role to cancel the double counted contributions. The $\Phi$ functional is the sum of 2PI graphs composed of dressed fermion propagators. We assume that a set of counter terms have been introduced to renormalize the vacuum self-energies and vertices. The number of counter terms can be either finite or infinite depending on the renormalizability of theories \cite{Weinberg:1978kz}, but this aspect is not important for our arguments. We also note that, because we write the action as a functional only for fermion propagators, the graphs we consider are not necessarily 2PI with respect to the propagators of other possible fields. 
 
The variation of $\Phi$ with respect to $S$ yields the graphs for the fermion self-energies, and if we choose $S$ to give the extrema, it satisfies the Schwinger-Dyson equation,
\beq
\frac{\, \delta I [S;\mu] \,}{\, \delta S (p) \,} 
= \tilde{\Sigma}^{ [S;\mu] } (p) - \Sigma^{[S]} (p) = 0 \,,~~~~~~~ \Sigma^{ [S] } (p) = - \frac{\, \delta \Phi [S] \,}{\, \delta S (p) \,} \,.
\eeq
The solution of the Schwinger-Dyson equation is written as $S^{\mu}_* (p)$. If we further differentiate $\Phi[S]$ by $S$, we get kernels with the 3-, 4-, and more-external legs. The equation of state at $\mu$ is obtained as 
\beq
 - P(\mu) =  I [S^{\mu}_*; \mu] - I [S^{\mu=0}_* ;\mu=0] \,,
\eeq
where the pressure is set to zero at $\mu=0$.

Now several comments are in order on the structure of the functional: 

(i) The expression of the 2PI functional differs from the expression of thermodynamic potential in the gaussian pair-fluctuation theories. The latter has the structure
\beq
I_{ {\rm GPF} } = I_{ {\rm MF} } + \Phi_2 [ S_{ {\rm MF} } ] \,,~~~~~~~
I_{ {\rm MF} } = \Tr_1 \Ln S_{ {\rm MF} } + \Tr_1 \left[ S_{ {\rm MF} }  \tilde{\Sigma}^{ [S_{ {\rm MF} } ;\mu] } \right] + \Phi_{ {\rm MF} } [S_{ {\rm MF} } ] \,,
\label{eq:functionalGPF}
\eeq
where $\Phi_2[S]$ is the sum of $\Phi_2^{ {\rm ring} } [S]$ and $\Phi_2^{ {\rm counter} } [S]$ defined in Eqs.(\ref{eq:gauss}) and (\ref{eq:counter}). The mean field part $I_{ {\rm MF} }$ is the 2PI functional and $\Phi_{ {\rm MF} }$ includes only 2PI graphs composed of a single fermion line. Here the solution of the Schwinger-Dyson equation in the mean-field level, $S_{ {\rm MF} }$, was substituted and $ \tilde{\Sigma}^{ [S_{ {\rm MF} } ;\mu] } = - \delta \Phi_{ {\rm MF} }/\delta S |_{ S_{ {\rm MF} } }$. We note that 
the structure of $I_{ {\rm MF} }$ has the form of the 2PI formalism. After just adding the gaussian fluctuations $\Phi_2 [S]$, however, we lose such correspondence by neglecting its impact on the fermion self-energy. In fact the incomplete treatment of such contributions introduces the double counting of some contributions; they are the origin of the UV divergences, as we will clarify shortly.

(ii) In the 2PI formalism the rernomalization of $n(>0)$-point functions in vacuum does not readily guarantee the UV finite equations of state. The situation is different from the 1PI effective action
$\Gamma[\phi]$, as a functional of some field values, $\phi$ \cite{Coleman:1973jx}. For the 1PI functional, we consider the form
\beq
\Gamma[\phi] - \Gamma[\phi=0] = \int_x \left[\, a (\partial \phi )^2 + b \phi^2 + c \phi^4 + \cdots \, \right]\,,
\label{eq:1PI}
\eeq
where $a, b, c, \cdots$ are some constants which must be renormalized. Here $a,b,c,\cdots$ appear in the $n$-point functions obtained from the functional derivative $\delta^n \Gamma/\delta \phi^n$, and they get renormalized through the studies of these functions. With the expression (\ref{eq:1PI}), the UV finiteness of $a, b, c, \cdots$ can be directly translated into the finiteness of the 1PI functional for a given distribution of $\phi(x)$. In contrast, the 2PI functional is characterized by a variable $S(x,y)$; even when the coefficients of $S$ are finite, this variable by itself can generate the divergence in the limit of $x\rightarrow y$. Therefore we need the discussions about the asymptotic behaviors of $S$.

\subsection{Isolating the divergence}

In the following we analyze the UV divergences in the functional. We first define
\beq
I_R [S;\mu] \equiv I[S ;\mu] - I[S^{\mu=0}_* ; \mu=0] \,,
\eeq
which is a functional of $S$ at finite $\mu$. To clarify the structure, it is convenient to distinguish the contributions associated with the change of fermion bases and the other, since the treatments of the UV divergences will be different for these two contributions. We decompose
\beq
I_R [S;\mu] = \left( I[S ;\mu] - I[S^\mu_\parallel ;\mu] \right) + \left( I[S^\mu_\parallel ;\mu] - I[S^{\mu=0}_* ; \mu=0] \right)
\equiv I_{\Delta S} [S;\mu] + I_{\Delta \mu}\,.
\label{eq:decomposition}
\eeq
Here the first term in the RHS, $I_{\Delta S} [S;\mu]$, measures the energy gain or cost associated with the change of bases, $S  \rightarrow S_\para$ (defined below), at the same chemical potential. Clearly at $S=S_\parallel^\mu$ the functional is  $I_{\Delta S} =0$, guaranteeing the existence of $S$ for which the functional $I_{\Delta S}$ is UV finite. On the other hand, the second term in Eq.(\ref{eq:decomposition}) compares the energies at different chemical potentials but with the same fermion bases; for this purpose we have introduced an in-medium propagator made of the vacuum fermion bases,
\beq
 S^\mu_{\parallel} (p) \equiv S^{\mu=0}_* ( p \rightarrow \tp) = \left[ - \Slash{\tp} + m + \Sigma^{\mu}_{\parallel} (p) \right]^{-1} \,, ~~~~~~  \Sigma^{\mu}_{\parallel} (p) \equiv \Sigma^{ \left[S_*^{\mu=0} \right] } (p \rightarrow \tp) \,.
\eeq
That is, we changed only the variable as $p_0 \rightarrow p_0-\rmi \mu$ but fixed all the others in the vacuum propagator. We emphasize that, except for the $\mu=0$ case, $S^\mu_\parallel$ in general does not minimize $I[S;\mu]$, so is not the solution of the Schwinger-Dyson equation, i.e., $\Sigma^{\mu}_{\parallel} \neq \Sigma^{[S^\mu_\para;\mu]} = - \delta \Phi/\delta S|_{S_\para^\mu}$. More about $S^\mu_\parallel$ will be discussed in the next section.

\begin{figure*}[!t]
\begin{center}
\vspace{-2.5cm}
\hspace{-.3cm}
\includegraphics[width = 0.8\textwidth]{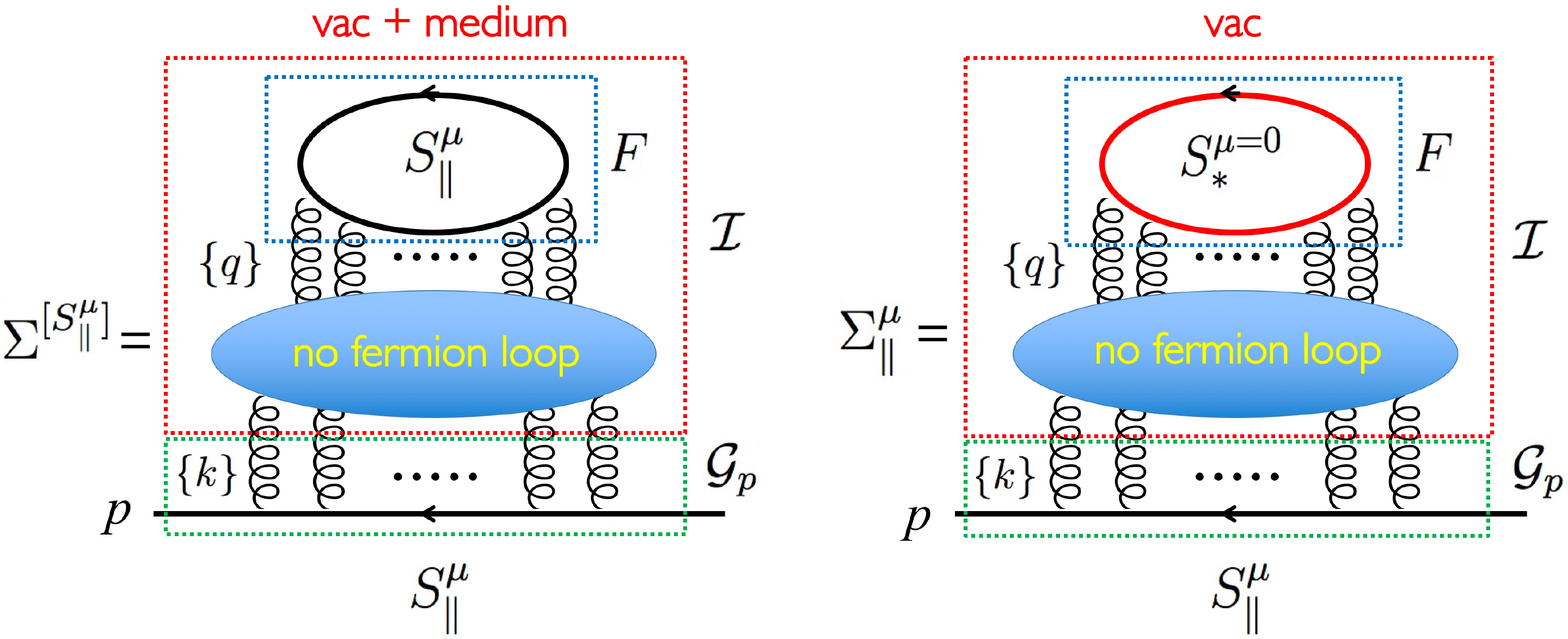}
\vspace{-0.cm}
\end{center}
\vspace{-2.5cm}
\caption{ 
\footnotesize{
The self-energies $\Sigma^{ [S^\mu_\para] } (p)$ and $\Sigma^\mu_{\para} (p)$ (the 1-fermion loop case as an illustration). In the former all fermion lines are calculated with the propagator $S^\mu_\para$, while in the latter all internal fermion loops are calculated with the vacuum propagator $S_*^{\mu=0}$. The fermion lines connected to the external legs are common for $\Sigma^{ [S^\mu_\para] } (p)$ and $\Sigma^\mu_{\para} (p)$; if no-fermion loop is available, these two self-energies coincide.}
\vspace{-0.0cm} }
\label{fig:dif_self_energy}
\end{figure*}

Below we will first give some concrete discussions for the in-medium propagators made of the vacuum fermion bases. After this preparation we discuss how $I_{\Delta \mu}$ and $I_{\Delta S}$ are made UV finite. We emphasize that the functional $I_{\Delta S} [S;\mu]$ is UV finite only for particular classes of $S$; the in-medium self-energy of $S$ must approach to the vacuum counterpart sufficiently fast at large momenta.

\section{In-medium calculations within the vacuum fermion bases}
\label{sec:same_bases}

\subsection{In-medium propagators made of the vacuum fermion bases}
\label{sec:in-meidum_propagator}

We first give more remarks concerning with the terminology ``fermion bases". We define them directly from the fermion propagators, rather than the unitary transformation of fermion field operators as the latter need not be manifestly treated. 
For the propagator $S^\mu_{\parallel} (p)$, most generally the spectral functions are the same as the vacuum one, and in this sense the propagator $S^\mu_{\parallel} (p)$ is made of the vacuum fermion bases. In the spectral representation,
\beq
 S^\mu_{\parallel} (p)
= - \int_0^\infty \! \frac{\, \rmd w \,}{\, 2\pi \,} \left[ 
\frac{\, \rho_\rmpp^{ {\rm vac} } (w, \vp ) \,}{\, \rmi p_0 + \mu -w  \,}
+ \frac{\, \rho_\rma^{ {\rm vac} } (w, \vp ) \,}{\, \rmi p_0 + \mu + w  \,} 
\right] \,,
\eeq
where $\rho_\rmpp^{ {\rm vac} }$, $\rho_\rma^{ {\rm vac} }$ are the spectral functions in vacuum ($4\times 4$ matrices) for particle and antiparticle components, respectively. Or one can also write the invariant mass parameterization
\beq
S^\mu_{\parallel} (p)
= - \int_0^\infty \! \frac{\, \rmd s \,}{\, 2\pi \,} \, \frac{\, \Slash{\tp} \rho_V^{ {\rm vac} } (s ) +  \rho_S^{ {\rm vac} } (s ) \,}{\, \tp^2 - s  \,}
\eeq
where  the Lorentz invariance in vacuum ensures that $\rho^{ {\rm vac} }_{V,S} (s)$ are the Lorentz scalar which depends only on the Lorentz scalar variable $s$. 
The vacuum spectral functions reflect that we are using the same fermion bases as in vacuum, while the modification $H\rightarrow H-\mu N$ ($H$: hamiltonian, $N$: number of fermions) is reflected only through the change of variable $p_0 \rightarrow p_0 - \rmi \mu$. 

\subsection{The self-energy} \label{sec:self_same_bases}

Below we compare the structure of $\Sigma^{[S_\para^\mu]} $ and $ \Sigma_\para^\mu$, then discuss both can be UV finite by the vacuum counter terms. 

These self-energies differ only when the fermion self-energy graph includes fermion loops, see Fig.\ref{fig:dif_self_energy}. 
The key observation is that the shift $p \rightarrow \tp$ in $ \Sigma_\para^\mu$ affects the propagator connected to the external leg, but does not affect the momenta of fermion loops; the fermion propagators in fermion loops are the vacuum one, $S_*^{\mu=0} ( = S_\para^{\mu =0} )$, that does not depend on $\mu$. On the other hand, in $\Sigma^{[S_\para^\mu]} $, all propagators, from the one connected to the external fermion lines as well as those forming loops,
are $S_\para^\mu$. So the formal structure of the difference in the self-energies can be written as
\beq
 \Sigma^{[S_\para^\mu]} (p) -  \Sigma_\para^\mu (p)
 = \int_{ \{k \} } \calG_p^{ [ S_\para^\mu ]} ( \{ k \} ) ~ 
 \left(\, \mathcal{I}^{ [ S_\para^\mu ] } ( \{ k \} )   - \mathcal{I}^{ [ S_\para^{\mu =0} ] } ( \{ k \} ) \, \right) \,,
 \label{eq:self-energy_general}
\eeq
where $\calG_p$ is the fermion line connected to the external fermion legs with momentum $p$, and the function $\mathcal{I}$ is the diagram attached to $\calG_p$. They are connected by lines carrying a set of momenta, $\{k \} = (k_1, k_2 , \cdots) $, which will be integrated. The difference in the self-energies is summarized in $\calI$ through the difference of fermion propagators in the fermion loops. From the definition it should be clear that the two self-energies coincide if there are no fermion loops.

\begin{figure*}[!t]
\begin{center}
\vspace{-2.5cm}
\hspace{-.3cm}
\includegraphics[width = 0.8\textwidth]{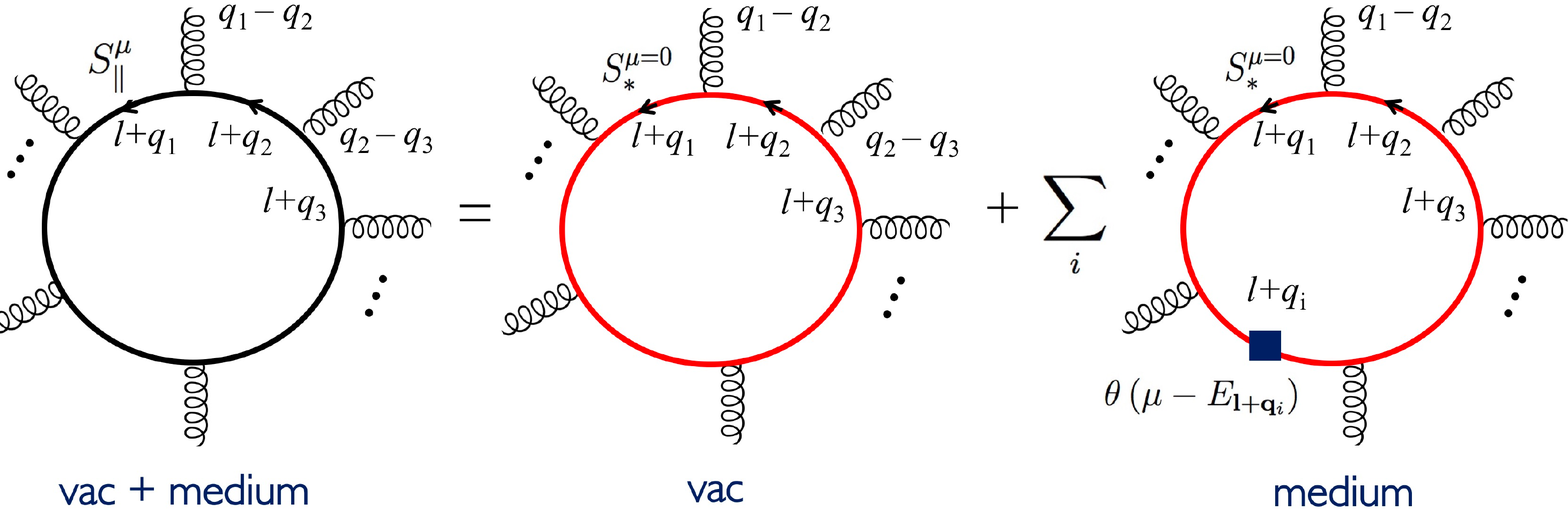}
\vspace{-0.cm}
\end{center}
\vspace{-2.5cm}
\caption{ 
\footnotesize{The 1-fermion loop graph with many line insertions, see the expression Eq.(\ref{formal_F}). The loop momentum is $l$. Using the standard technique of the analytic continuation one can isolate the vacuum piece and the medium dependent piece, as shown in Eq.(\ref{formal_F2}). The box in the last diagram specifies where we pick up the pole.
}
\vspace{-0.0cm} }
\label{fig:loop_residue}
\end{figure*}

\subsubsection{Fermion loops} \label{subsec:a fermion loop}
To see how $\mathcal{I}^{ [ S_\para^\mu ] } ( \{ k \} )$  and $ \mathcal{I}^{ [ S_\para^{\mu =0} ] } ( \{ k \} ) $ are calculated, we focus on the subdiagrams in which there is one fermion loop and external lines with momenta $q_1, q_2 ,\cdots$ are attached, see Fig.\ref{fig:loop_residue}. Let $l$ be the loop momenta. Then all fermion propagators $S_\para$ in the loop has the $\mu$-dependence only through the combination of $\tl_0 = l_0 - \rmi \mu$. More explicitly such a loop contains the structure (reminder: $S_\para^\mu (p) = S_*^{\mu =0 }(p\rightarrow \tp)$)
\beq
F^{ \{ q\} }_{\vl} ( \tl_0 ) =  -\tr \left[ S^{\mu=0}_* ( \tl +q_1 ) V_{1,2} S^{\mu=0}_* ( \tl + q_2 ) V_{2,3}  S^{\mu=0}_* ( \tl + q_3)  \cdots \right] \,,
\label{formal_F}
\eeq
in which several external lines with momenta $q_{i=1,2,\cdots}$ are attached to the fermion loop with the vertices $V$.  For the moment we introduce an infinitesimal temperature $T$ and use the Matsubara formalism with $\tl_0 \rightarrow \omega_n - \rmi \mu$ where $\omega_n = (2n+1)\pi T$ \cite{Kapusta}. 
Here $\mu$ is hidden in the variable $\tl_0$, so the integral over $l_0$ leads to
\beq
\int_l F^{ \{ q\} }_{\vl} ( \tl_0 ) 
~\rightarrow~
T \sum_n \int_{\vl} F^{ \{ q\} }_{\vl} ( \omega_n -\rmi \mu )
= - \int_{\vl} \int_C \frac{\, \rmd l_0 \,}{ 2\pi \rmi } \frac{ F^{ \{ q\} }_{\vl}  ( -\rmi l_0 ) }{\, \rme^{\frac{\, l_0 -\mu \,}{T} }+ 1 \,}  
\,,
\label{formal_F2}
\eeq
where $C$ is the contour surrounding the poles of $( e^{ \frac{ l_0 -\mu \,}{T} }+ 1)^{-1}$, and now this factor is the only place where the $\mu$-dependence appears.
Next we pick up poles from each propagator by deforming the contour $C$ in the standard way. Here we consider the poles of the $i$-th propagator in the loop $F^{ \{ q\} }_{\vl} $. For this calculation the most general expression can be obtained by using the spectral representation,
\beq
S^{\mu=0}_* \left( -\rmi l_0 + q_{i0}, \vl+\vq_i \right)
= - \int_0^\infty \! \frac{\, \rmd w_i \,}{\, 2\pi \,} \left[ 
\frac{\, \rho_\rmpp^{ {\rm vac} } (w_i, \vl+\vq_i ) \,}{\, l_0-w_i + \rmi q_{i0} \,}
+ \frac{\, \rho_\rma^{ {\rm vac} } (w_i, \vl+\vq_i ) \,}{\, l_0 + w_i + \rmi q_{i0} \,} 
\right] \,,
\eeq
where $q_{i0}$ is left Euclidean.
After picking up the residues $l_0 = \pm w_i - \rmi q_{i0} $, the statistical factors in Eq.(\ref{formal_F2}) for particles become, in the zero temperature limit,
\beq
 \frac{ 1 }{\, \rme^{\frac{\, w_i -\mu - \rmi q_{i0}\,}{T} }+ 1 \,}
 ~\rightarrow ~  \theta \left( \mu - w_i \right) \,,
 \eeq
and for antiparticles
\beq
 \frac{ 1 }{\, \rme^{\frac{\, - w_i -\mu - \rmi q_{i0}\,}{T} }+ 1 \,}~\rightarrow ~  1\,.
\eeq
The antiparticle contributions are totally made of the vacuum quantities; after all the two loop graphs are expressed as the sum of the vacuum pieces and the medium dependent part,
\beq
\int_l  F^{ \{ q\} }_{\vl} ( \tl_0 ) 
= \int_l  F^{ \{ q\} }_{\vl} ( l_0 ) - \sum_i \int_0^\infty \rmd w_i  \int_{\vl}  \theta \left( \mu - w_i  \right) \tilde{F}^{ \{q ; i \} }_{\vl} (-\rmi w_i  ) \,,
\label{eq:loop_F_general}
\eeq
where ($l_w = (-\rmi w_i, \vl)$ and $q_{i,j} = q_i - q_j$)
\beq
  \tilde{F}^{ \{q ; i \} }_{\vl} (-\rmi w_i  )
  = -
 \tr \left[ \cdots S^{\mu=0}_* ( l_w + q_{i-1,i} ) V_{i-1, i } \left(\, \rho_\rmpp^{ {\rm vac} } (w_i, \vl ) \, \right) V_{ i, i+1 } S^{\mu=0}_* ( l_w +q_{i+1,i}  ) \cdots \right] \,.
\label{formal_F2}
\eeq
After the integration of $l_0$, the vacuum spectral density $\rho_\rmpp^{ {\rm vac} } $ remains as contributions from the $i$-th propagator.
Note that we have shifted variables $\vl+\vq_i$ in the spectral density so that the $j$-th propagator depends on $\{q\}$ through the combination $q_{j,i}$.

Now we can see the domain of the spatial momenta $\vl$ is bound and the integral over $\vl$ is UV finite. To confirm this we use the fact the energy of the vacuum spectral function behaves as $w = \sqrt{ m^2 + \vl^2}$ for a state with the invariant mass $m^2$. Then we notice that there is the energy bound $\theta(\mu-w)$ which in turn limits the domain of $\vl$. 
The result is entirely expressed by the vacuum spectral density and $\mu$, as we used the propagators $S_\para^{\mu}$ and $S_\para^{\mu=0}$.

\subsubsection{Integrating lines connected to fermion loops} \label{subsec:connected}
We have just verified the fermion loops in the subgraphs are divided into the vacuum pieces plus UV finite pieces. The last step for the evaluation of the self-energies is to integrate out momenta connecting the fermion loops to the external fermion lines. 

\begin{figure*}[!t]
\begin{center}
\vspace{-0.5cm}
\hspace{-.3cm}
\includegraphics[width = 0.4\textwidth]{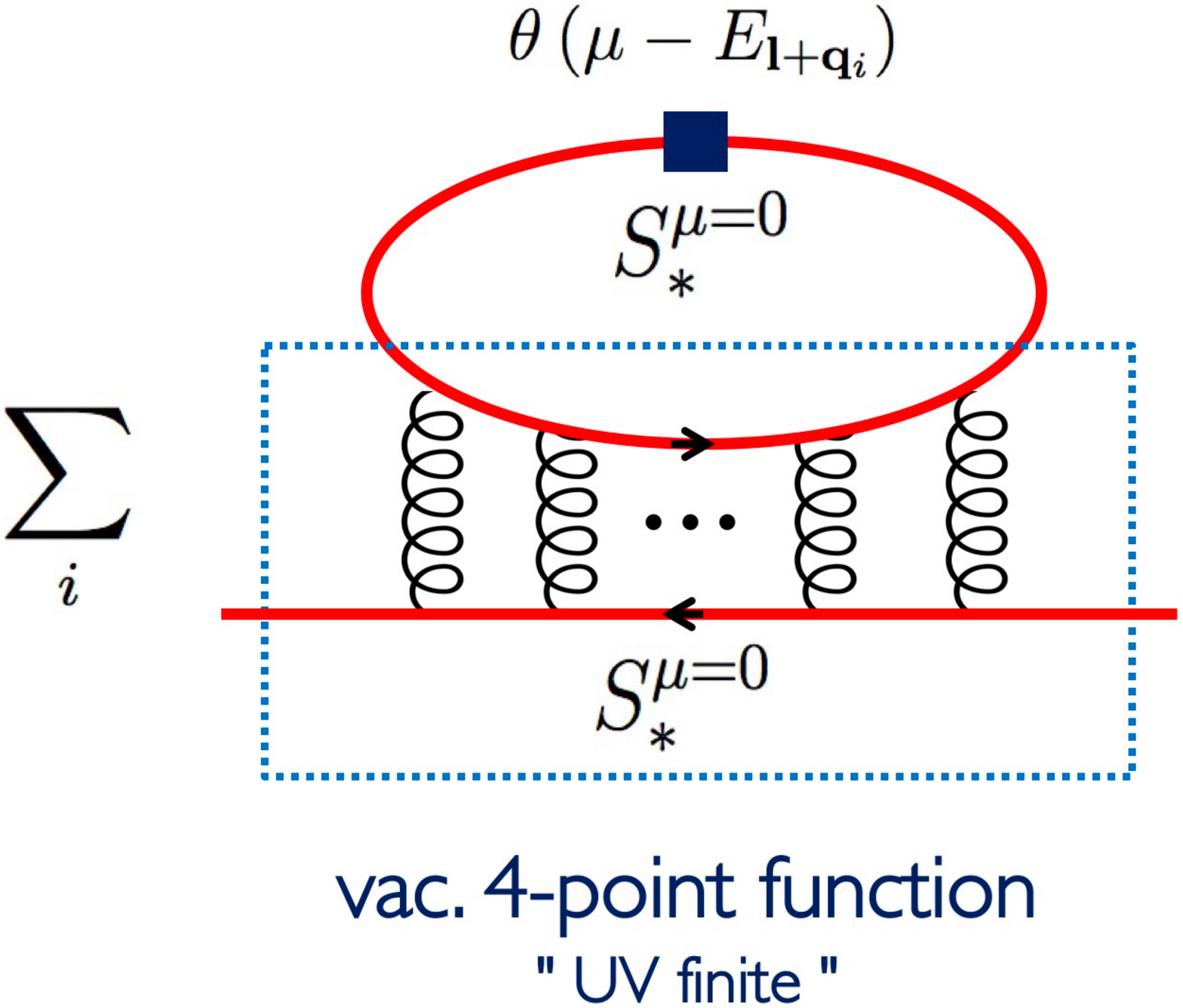}
\vspace{-0.cm}
\end{center}
\vspace{-0.5cm}
\caption{ 
\footnotesize{ The next-to-worst UV contribution with 1-medium insertion (the 1-fermion loop case as an illustration). The vacuum-medium piece is UV finite, as we can regard the graph originated from the contraction between $\theta$-function and the vacuum 4-point function which can be renormalized by the vacuum counter terms.
}
\vspace{-0.0cm} }
\label{fig:self_energy_NLO}
\end{figure*}

As an example a one-fermion loop graphs for the self-energy difference, $\Sigma^{[S_\para^\mu]} (p) -  \Sigma_\para^\mu (p)$, are shown in Fig.\ref{fig:self_energy_NLO}. The vacuum part in the fermion loop was already cancelled by taking the difference. Note that one of fermion lines in the loop is replaced with the vacuum spectral density with the restriction $\theta(\mu- E_{\vl+\vq_i } )$ attached, as in Eq.(\ref{formal_F2}). The graph in Fig.\ref{fig:self_energy_NLO} can be evaluated by using the vacuum 4-point functions which are renormalized and UV finite. The only question is whether closing two legs with momenta $\vl$ generates the UV divergence, but its integration domain is restricted, so no new UV divergences arise. So we have proved that $\Sigma^{[S_\para^\mu]} (p) -  \Sigma_\para^\mu (p)$ is UV finite. Moreover $ \Sigma_\para^\mu (p)$ and $\Sigma^{[S_\para^\mu]} (p) $ are separately UV finite; the former is the vacuum self-energy with replacement $p\rightarrow \tp$ and hence is UV finite; thus $\Sigma^{[S_\para^\mu]} (p) $ is also UV finite. 

The case with more fermion loops does not introduce any essential modifications. Essentially such graphs can be regarded the vacuum $n$-point functions contracted to the medium $\theta$-functions. The vacuum functions are assumed to be UV finite, and the contraction only leads to integrals whose domains are limited, so the results of the integration are all finite.

We note that the present argument does not ask whether the theory is renormalizable or not. If all vacuum functions are made UV finite for some fermion bases, and if we keep using the same bases from the vacuum to a medium, then the in-medium self-energies are UV finite. 
The key fact was that the UV contributions {\it exactly} cancel in Eq.(\ref{eq:loop_F_general}). If we used different bases for the vacuum and medium, the cancellation was not exact; in such cases we must consider the power counting for asymptotic momenta, as we will discuss in the next section. 

\subsection{The UV finiteness of $I_{\Delta \mu}$ }

\begin{figure*}[!t]
\begin{center}
\vspace{-2.5cm}
\hspace{-.3cm}
\includegraphics[width = 0.8\textwidth]{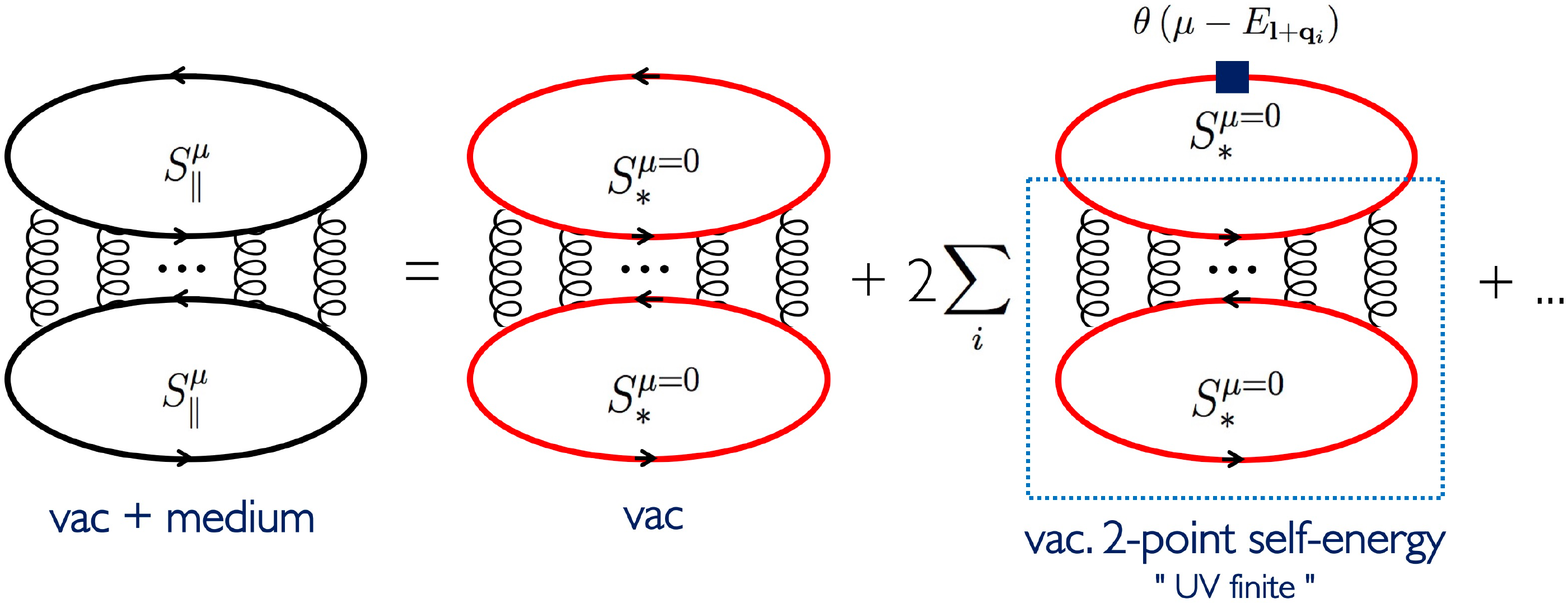}
\vspace{-0.cm}
\end{center}
\vspace{-2.5cm}
\caption{ 
\footnotesize{An example of the $\Phi$-functional with 2-fermion loops. In each loop we decomposed the vacuum and medium dependent parts, and then considered the product. The vacuum-vacuum product has the UV divergence that can be subtracted off. The 1-medium insertion graphs are UV finite, as we can regard the graph originated from the contraction between $\theta$-function and the renormalized vacuum 2-point self-energy. The same logic is applied for higher orders of medium insertions.
}
\vspace{-0.0cm} }
\label{fig:zero_point_function}
\end{figure*}

Finally we consider $I_{\Delta \mu}$. The naive power counting of the dimensionality indicates the presence of terms like $\sim \mu^2 \Lambda_{ {\rm UV}}^2$ or $\sim \mu^4 \ln \Lambda_{ {\rm UV} }$. The vacuum subtraction, however, completely cancels these UV divergences as far as we use the same fermion bases for the vacuum and medium. This is what happens in perturbation theory and has been discussed in the standard textbook \cite{Kapusta}. For the later purpose we briefly review this cancellation in somewhat abstract fashion, looking at general graphs which may include infinite number of loops. 

The discussion goes in the very similar way as one given for the self-energy, see Fig.\ref{fig:zero_point_function} for the zero-point energy.
There may be several fermion loops. But as we have already discussed, we can decompose each loop into the vacuum term and medium part with $\theta$-function inserted.
Thus by considering all fermion loops we find the product of the vacuum term and the medium terms.

The potentially worst UV divergence would appear if we pick up the vacuum contributions from all fermion loops, but it is the vacuum quantity and can be cancelled by the vacuum subtraction. The next-to-worst divergence appears if we pick up only one medium piece but choose the vacuum pieces for all the rest of fermion loops. But such diagrams can be regarded as vacuum fermion 2-point functions contracted with the medium piece. The vacuum 2-point function is UV finite by our assumption on the vacuum renormalization, and its contraction of the external fermion legs with the medium dependent piece, bound by $\mu$, does not yield any additional UV divergence. Thus the 2PI graphs with one medium piece are UV finite. The graphs with more medium pieces can be discussed in the same way. With this we conclude $I_{\Delta \mu}$ is UV finite.

\section{UV contributions induced by changes of fermion bases}
\label{sec:change_bases}

\subsection{The UV finiteness of $I_{\Delta S}$ and constraints on the fermion bases}
\label{sec:change_bases_n_point}

We have already seen that the 2PI functional in medium is finite if we keep using the same fermion bases as in the vacuum. But the medium effects often deform the fermion bases, typically in a non-perturbative manner. The purpose here is to examine that to what extent we can change the fermion bases without encountering the UV problems. 
If we met UV divergences, we would be forced to keep the bases same as the vacuum one (for which $I_{\Delta S}$ is guaranteed to be zero). It is important to emphasize that, unless correctly handling the UV contributions, we might artificially exclude physical solutions.

For these reasons we analyze in detail the UV structure of the functional $I_{\Delta S}$ (defined in Eq.(\ref{eq:decomposition})) and the condition required for $S$. We classify the strength of the UV contribution by choosing the expansion parameter as (reminder: $\tilde{\Sigma}^{ [S;\mu] } = S^{-1}- ( S^{\mu}_{ {\rm tree} } )^{-1}$ 
which becomes the physical self-energy when we choose $S$ to be the solution of the Schwinger-Dyson equation)
\beq
S^{-1} (p) - S^{-1}_\para (p) = \tilde{\Sigma}^{ [S;\mu] } (p) - \Sigma^{\mu}_{\parallel} (p) \equiv \Delta \Sigma^{[S;\mu]} (p) \,.
\eeq
This measures the difference between $S$ and $S_\para$. We count $\tilde{\Sigma}^{ [S;\mu] } \sim \Sigma^{\mu}_{\parallel} \sim p$ at large $p$. For the physical solution of $S$, we expect that at large $p$ the IR effects decouple; if this is the case $\Delta \Sigma \rightarrow 0$ as $p\rightarrow \infty$. The question is how fast this damping is. To characterize the damping, we write
\beq
 \Delta \Sigma \sim \tp^{\,-1 + \gamma} \,.
\eeq
with which 
\beq
S (p) - S_\para (p) \sim  - S_\para (p)  \Delta \Sigma  S_\para (p) \sim \tp^{\,-3 + \gamma} \,,
\eeq
modulo possible logarithmic factors in $p$.
Here we consider the power of $\tp$, not $p$, because the counting based on the latter involves the expansion of $\mu/p$ which is not helpful. We characterize $S$ by this damping. 
The asymptotic behaviors of $S - S_\para $ are classified by the value of $\gamma$: 

(a) The $\gamma=0$ case will be called `canonical' and we regard it as our baseline; powers are reduced by 2 compared to the single fermion propagators, $S - S_\para \sim \tp^{-3}$; 

(b) The $\gamma=-1$ case happens when we omit the vector self-energies and include only mass self-energies, leading to $S - S_\para \sim \tp^{-4}$; 

(c) The $\gamma =1$ case happens if we include only mass self-energies as in (b) but assume that the medium-induced constant shift of the self-energies, e.g., $\Delta \Sigma \sim \Delta M= $ const, survives to large momenta, leading to $S - S_\para \sim \tp^{-2}$; 

(d) The $\gamma=2$ case happens if the medium-induced shift in the residue function survives to large momenta and couples to the $\Slash{p}$-component, e.g., $\Delta \Sigma \sim \Delta Z \Slash{p}$, with $\Delta Z=$ const, leading to $S - S_\para \sim \tp^{-1}$; 

(e) The case $\gamma >2$ is regarded unlikely, so this case will be excluded from our considerations, unless otherwise stated.

In the following, models in our mind are those of Yukawa or gauge theory types in which fermion couples to bosons whose propagators behave as $\sim p^{-2}$ in the UV domain. To simplify the presentation we assume the absence of tadpole-type graphs. If there are tadpoles, the self-energies acquire a constant independent of the external momenta, as in the cases (c) and (d) which will require extra discussions of shifting the (bosonic) field, $\phi \rightarrow \phi_0 + \phi$. Such procedures will not be performed in this paper.

\subsection{Preparation: power counting for the subgraphs}

The discussion of the UV behaviors get involved because of loops in subgraphs. To disentangle various UV divergences shared by different loops, it is useful to specify the soft and hard parts in the graphs. We assign powers for fermion propagators as
\beq
S \sim \left\{ \begin{matrix}
&p^{-1} ~~~~&( {\rm hard} ) \\
& \calM^{-1}  ~~~~&( {\rm soft} )
\end{matrix}\right.
\eeq
where $\calM$ is a constant with the mass dimension one. Accordingly the fermion self-energy is
\beq
\Sigma \sim \left\{ \begin{matrix}
&p ~~~~&( {\rm hard} ) \\
& \calM  ~~~~&( {\rm soft} )
\end{matrix}\right.
\eeq
For boson propagators we assign
\beq
D \sim \left\{ \begin{matrix}
&p^{-2} ~~~~&( {\rm hard} ) \\
& \calM^{-2}  ~~~~&( {\rm soft} )
\end{matrix}\right.
\eeq
Finally we also assign powers for loop integration,
\beq
\int_l ~~\sim \left\{ \begin{matrix}
&p^{4} ~~~~&( {\rm hard} ) \\
& \calM^{4}  ~~~~&( {\rm soft} )
\end{matrix}\right.
\eeq
where the domain of integration is restricted to the IR region for the soft part.

\begin{figure*}[!t]
\begin{center}
\vspace{-2.5cm}
\hspace{-.3cm}
\includegraphics[width = 0.9\textwidth]{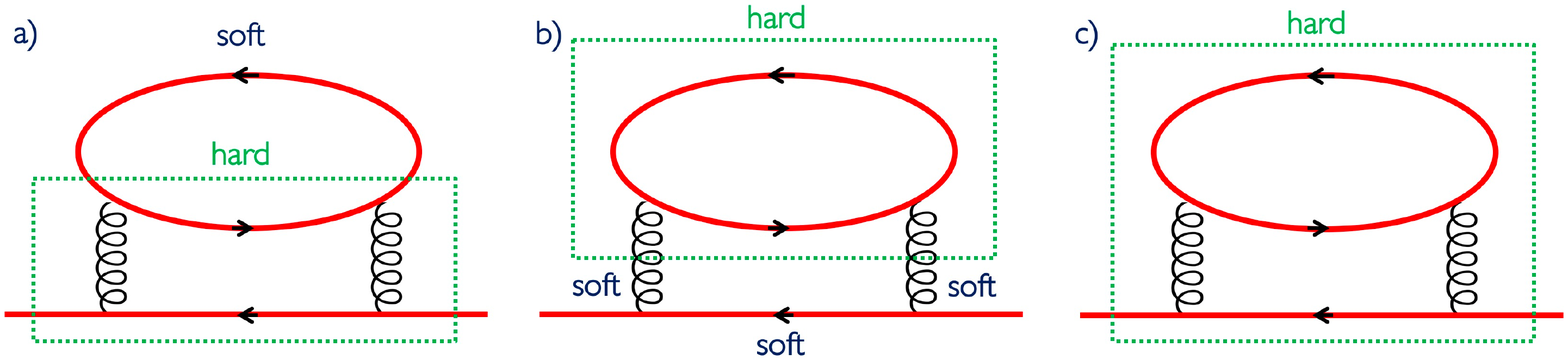}
\vspace{-0.cm}
\end{center}
\vspace{-3.2cm}
\caption{ 
\footnotesize{An example of 2-loop graph. In a box, lines participating in a loop carry hard momenta. 
}
\vspace{-0.0cm} }
\label{fig:soft-hard}
\end{figure*}

As an illustration we consider a 2-loop graph for the fermion self-energies where one-fermion loop is connected to external lines by 2-boson lines (Fig.\ref{fig:soft-hard}). We use a box to indicate that inside the box the lines participating in the loops are hard. 

For the graph (a), one fermion line in the loop is kept soft but all the other lines are set hard. Then the counting is
\beq
\calM^4 p^4 \times \left( p^{-1}  p^{-4}  p^{-1} \calM^{-1} \right) \sim \calM^3 p^{-1} \,,
\eeq
where the first is the factor from the phase space; $\calM^4$ is the phase space for the soft fermion line and $p^4$ for the other loop momenta. The factor in the parenthesis is from propagators. Therefore this case (a) is $\sim p^{-1}$ and hence UV finite for $p\rightarrow \infty$.

For the graph (b), the fermions in the loop are hard but the others are all soft. The power counting is
\beq
\calM^4 p^4  \times \left( \calM^{-1} \calM^{-4} \times p^{-2} \right) \sim p^2 \,.
\eeq
For the graph (c), all lines are hard, so 
\beq
p^8  \times \left( p^{-1} p^{-4} \times p^{-2} \right) \sim p^2 \,.
\eeq
The graphs (b) and (c) are the order of $\sim p^2$ and apparently have the quadratic divergence whose origin is the hard fermion loop. The leading quadratic divergence is cancelled by the vacuum subtraction, but generally that procedure still leaves the logarithmic divergence coupled to quantities dependent on the fermion bases. To eliminate such contributions we need special cares by summing a proper set of graphs. We will come back to this point when we discuss the boson self-energy in Sec.\ref{sec:boson-self}.

\subsection{The self-energies for fermions}\label{sec:fermion-self}

Before looking at the zero-point energy, we compare the self-energies for different bases and examine how the difference can be made finite. More explicitly we will assume $\gamma \sim 0$ for the counting and then check that the self-energy graphs are finite under this assumption; if not we would run into inconsistencies. Here it is sufficient to discuss the $\mu=0$ case for the divergences associated with the change of bases; as we have already seen, the in-medium and vacuum self-energies for the same bases differ only by the UV finite value.

The structure of fermion self-energies looks relatively simple because from the dimensional ground they apparently have only the logarithmic divergences. This is indeed the case if there are no fermion loops inside of the self-energy graphs. Extra cares are necessary for graphs with internal fermion loops in the subgraphs. To see it we consider the difference in the self-energies for propagators $S^{\mu=0}$ and $S_*^{\mu=0}$,
\beq
\left( \Sigma^{ [S^{\mu=0} ] } - \Sigma^{ [S_*^{\mu=0}] } \right)_{\alpha \beta} 
= \frac{\, \delta \Sigma_{\alpha \beta} \,}{\delta S_{\gamma \delta}} \bigg|_{S^{\mu=0}_*} \left( S^{\mu=0} - S^{\mu=0}_* \right)_{\gamma \delta} + \cdots
\eeq
where the first term in the RHS represents the sum of graphs in which one fermion line is set to $S^{\mu=0} - S^{\mu=0}_* $ and all the others are $S^{\mu=0}_*$.

\begin{figure*}[!t]
\begin{center}
\vspace{-0.5cm}
\hspace{-.3cm}
\includegraphics[width = 0.5\textwidth]{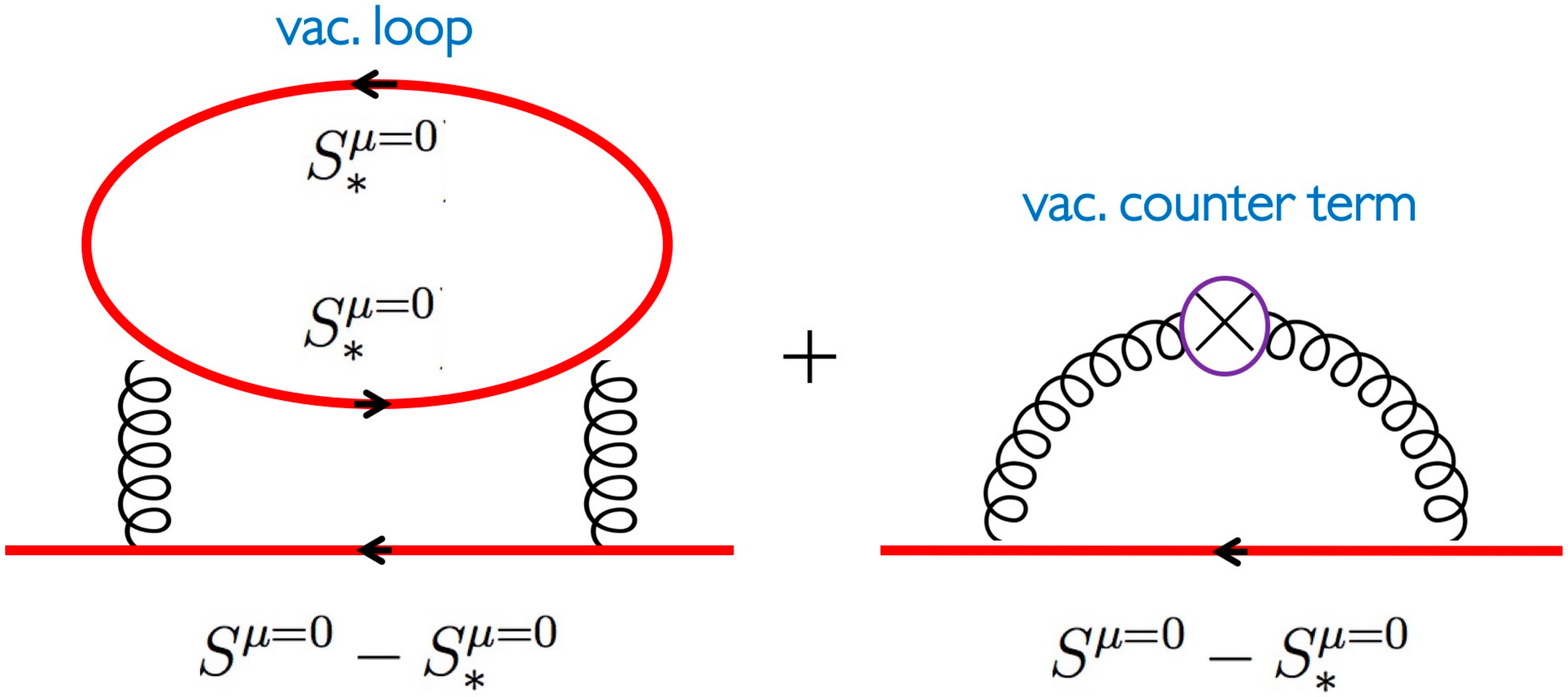}
\vspace{-0.cm}
\end{center}
\vspace{-1.2cm}
\caption{ 
\footnotesize{A graph generated from the expansion of $\Sigma^{ [S^{\mu=0} ] } - \Sigma^{ [S_*^{\mu=0}] }$ (with 2-boson lines) to the linear order of $S^{\mu=0} - S^{\mu=0}_* $. The divergence from the vacuum fermion loop is cancelled by the counter term in the second graph, to leave powers $\sim l^2$. It is combined with the boson propagators and a subtracted fermion propagator, leading to the UV finite result. 
}
\vspace{-0.0cm} }
\label{fig:2loop_s-s_type1}
\end{figure*}

\begin{figure*}[!t]
\begin{center}
\vspace{-0.5cm}
\hspace{-.3cm}
\includegraphics[width = 0.3\textwidth]{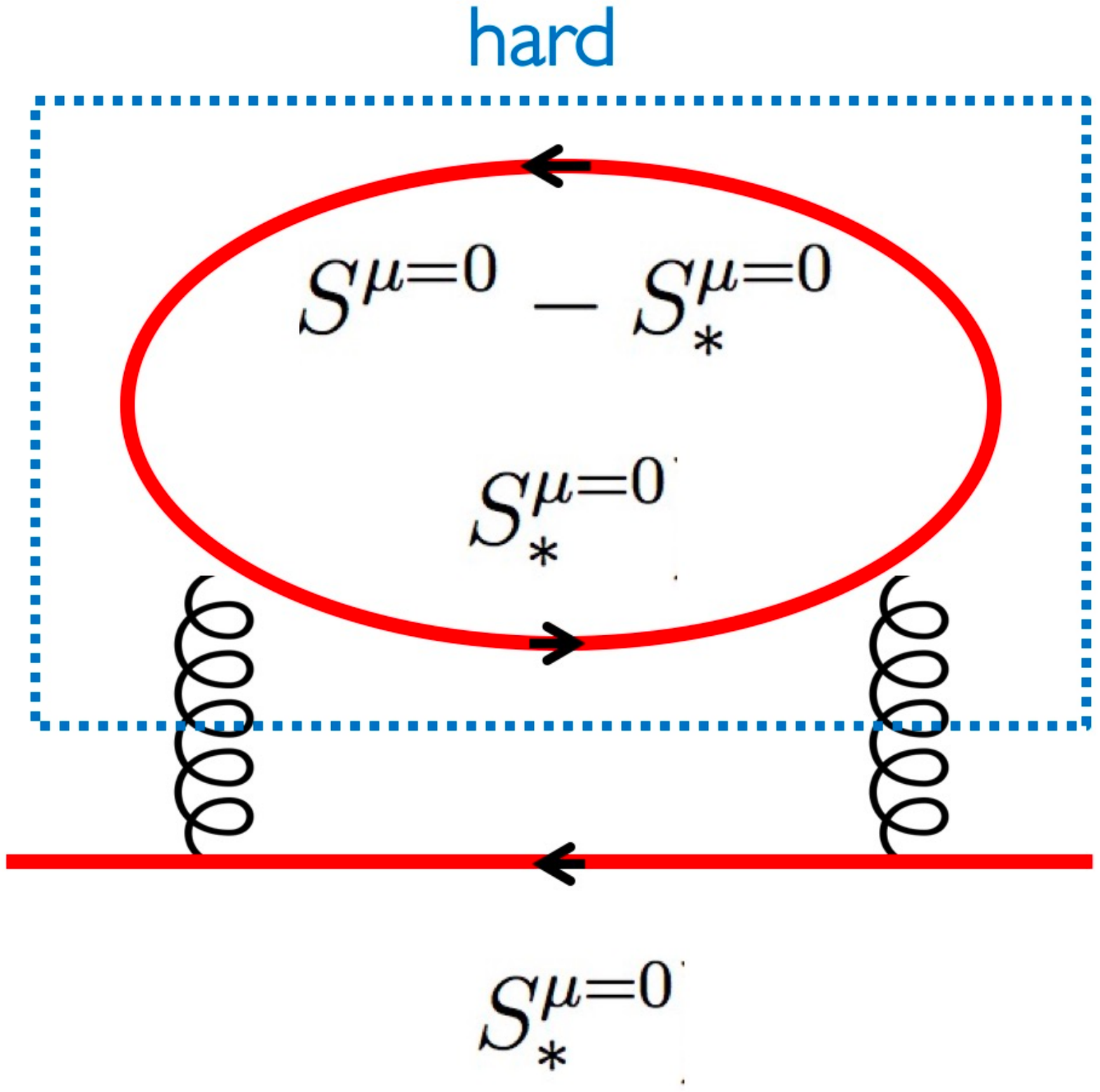}
\vspace{-0.cm}
\end{center}
\vspace{-2.2cm}
\caption{ 
\footnotesize{A graph generated from the expansion of $\Sigma^{ [S^{\mu=0} ] } - \Sigma^{ [S_*^{\mu=0}] }$ (with 2-boson lines) to the linear order of $S^{\mu=0} - S^{\mu=0}_* $. The bubble graph contains a single subtracted propagator $S^{\mu=0} - S^{\mu=0}_* $. Naive counting leads to $\sim \ln \Lambda_{ {\rm UV} }$ in the fermion loop. To eliminate it other graphs not shown here (with more boson line insertions) must be added. The way of the cancellation depends on models.}
\vspace{-0.0cm} }
\label{fig:2loop_s-s_type2}
\end{figure*}

As an example, we consider a 2-loop graph shown in Figs.\ref{fig:2loop_s-s_type1} and \ref{fig:2loop_s-s_type2}.
The above expansion produces two types of graphs. In the first type (the graph in Fig.\ref{fig:2loop_s-s_type1}), a subtracted propagator $S^{\mu=0} - S^{\mu=0}_* $ belongs to the line shared with the external legs.
In the second type (the graph in Fig.\ref{fig:2loop_s-s_type2}), one line in the fermion loop is set to $S^{\mu=0} - S^{\mu=0}_* $.
 
In the first type, a momentum $l$ enters the fermion loop made of the vacuum propagators. As we saw in the last section, the UV divergence would happen only in the case where hard momentum enters the fermion loop. The loop is made of the vacuum bases, so with our assumption on the vacuum renormalization, it is made UV finite by the counter term (or some cutoff). Then, from the dimensionality the loop leaves powers $\sim l^2$ (modulo logarithm) with UV finite coefficients. Now this piece is contracted with two gluon propagators attached to the fermion line, to yield factors $ l^2 \times (l^{-2} )^2 \times l^{-3} \sim l^{-5}$ (factors from two boson propagators; the internal fermion loop; and $S^{\mu=0} - S^{\mu=0}_* $). Thus the integral over $l$ leaves UV finite functions of fermion external momenta. 

Next we consider the graph in Fig.\ref{fig:2loop_s-s_type2}, where one line in the fermion loop is set to $S^{\mu=0} - S^{\mu=0}_* $. Again it is sufficient to consider the divergence from the fermion loop. The loop include a subtracted fermion propagator of $\sim l^{-3}$, but this reduction of powers is not sufficient to make the loop convergent; there would remain the logarithmic divergence. No vacuum counter term is available to eliminate this divergence. Hence this divergence must be cancelled by assembling the divergences from other graphs. This divergence in the fermion loop also propagate to the fermion self-energies.

Before discussing how to handle this divergence, it is more convenient to rephrase it in terms of boson self-energies. Let $\Pi^{[S]}$ a correlator of quark bilinear currents. This is the fermion contributions to the boson self-energies. We note an algebraic relation,
\beq
\Pi^{ [S^{\mu=0} ] } = \Pi^{ [S_*^{\mu=0} ] } - 2  \Tr[ S_*^{\mu=0}V(S^{\mu=0}-S^{\mu=0}_* )V ] - \Tr[ (S^{\mu=0} - S^{\mu=0}_* ) V (S^{\mu=0}-S^{\mu=0}_*)V ] \,,
\eeq
where the second term in the RHS is what we are discussing. The first term in the RHS is the vacuum functions and can be made finite by the vacuum counter term, while the integral of the third term is convergent and is the order of $l^{-2}$. Therefore proving the UV finiteness of the second term in the RHS is equivalent to proving the UV finiteness of the LHS. 

Actually the UV finiteness of $ \Pi^{ [S^{\mu=0} ] } $ is necessary but not sufficient condition to make the fermion self-energy finite. From the dimensional ground $ \Pi^{ [S^{\mu=0} ] } \sim l^2$, but we need $\Pi^{ [S^{\mu=0} ] } - \Pi^{ [S_*^{\mu=0} ] } \sim l^0$, i.e., the leading contributions of $l^2$ to cancel. Otherwise the self-energy difference of $\sim l^2$ couple to two boson propagators to yield $l^{-2}$. Then the fermion self-energy may have the logarithmic divergence. 

From these arguments, now our problem about the fermion self-energies is now translated to the problem of proving the boson self-energies to be UV finite and $\sim l^0$.

\begin{figure*}[!t]
\begin{center}
\vspace{-.5cm}
\hspace{-.3cm}
\includegraphics[width = 0.5\textwidth]{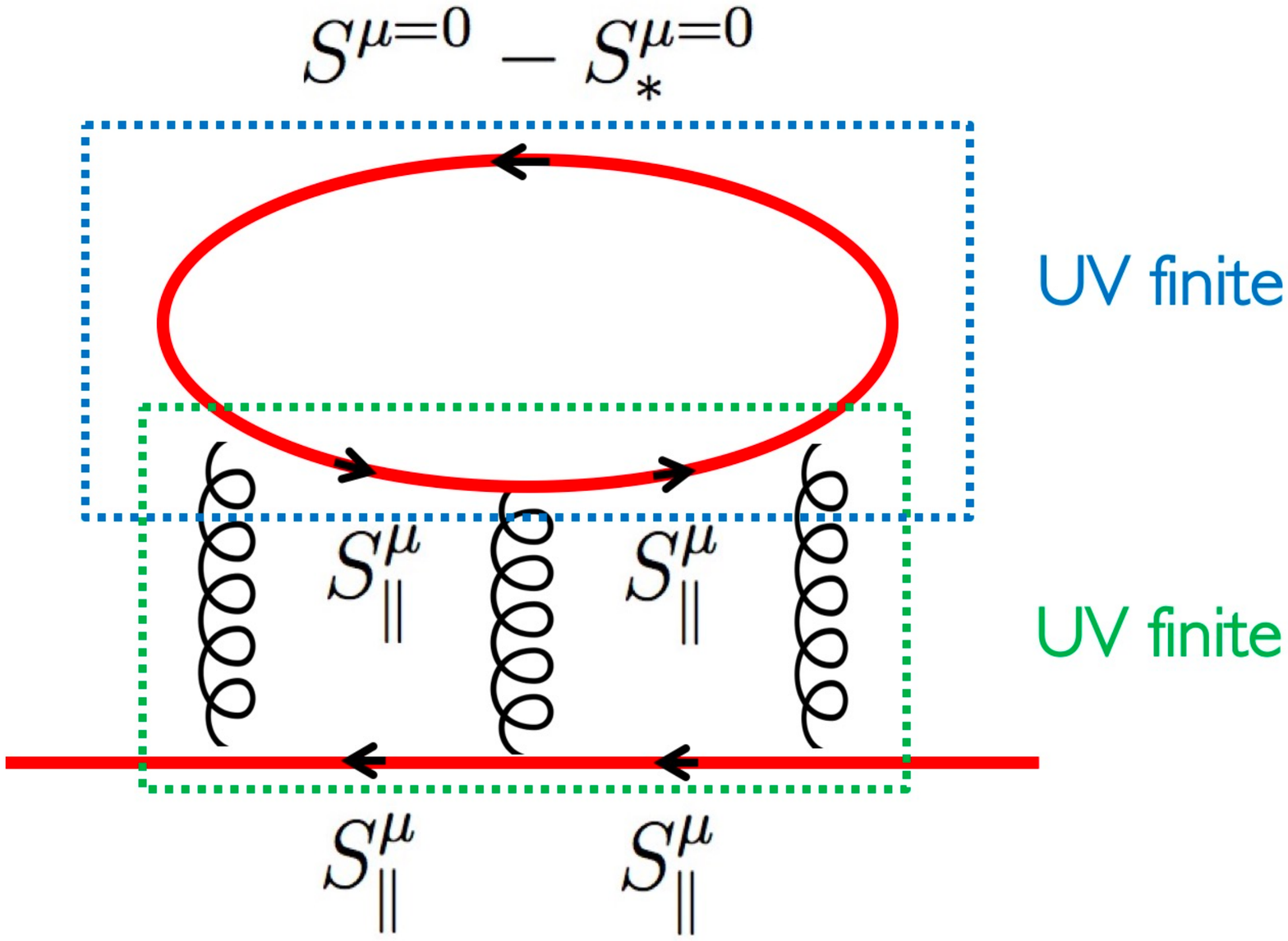}
\vspace{-0.cm}
\end{center}
\vspace{-1.cm}
\caption{ 
\footnotesize{A graph contributing to $\Sigma^{ [S^{\mu=0} ] } - \Sigma^{ [S_*^{\mu=0}] }$ with 3-boson lines.  The linear order of $S^{\mu=0} - S^{\mu=0}_* $ appear in the fermion loop. This subgraph is UV finite as each boson line insertion reduces the power by one. The other subgraphs made of 4-point fermion functions are also UV finite.}
\vspace{-0.0cm} }
\label{fig:3loop_examplei}
\end{figure*}

Before proceeding to the boson self-energies, we mention what happens if there are more boson lines attached to the fermion loop and to the fermion line connected to the external legs. An example is shown in Fig.\ref{fig:3loop_examplei}. In this case the analyses are actually simpler than the case shown in Fig.\ref{fig:2loop_s-s_type2}; the fermion loop more than two boson line insertions is by itself UV convergent, so that there is no danger to get the UV divergent factor from this subgraph. Also, the 2-loop subgraph with 3-boson lines can be regarded as the vacuum 4-point fermion functions is convergent by itself. So if one of the loop is soft, the result is convergent. Finally, when all loops are hard, the 4-point function is $\sim l^{-2}$ and $S^{\mu=0} - S^{\mu=0}_* \sim l^{-3}$, so the integration over $l$ leaves UV finite quantities.  Therefore subgraphs, in which the fermion loop with only two boson line insertions, is exceptional and requires special cares.

\subsection{The self-energies for bosons}\label{sec:boson-self}

The discussion for the bosonic self-energies should be much more involved than the fermion cases, since the naive power counting indicates its leading divergence to be the quadratic order. The leading divergence can be cancelled in the difference of the self-energies, while the next-to-leading order UV divergence (logarithmic divergence) may couple to the quantities dependent on the fermion bases. Our task is to identify the recipe to remove the logarithmic divergence. 

Such recipe requires the detailed discussions which depend on models. There have been detailed analyses in the literatures which will not be repeated here; we give only the outline of the necessary discussion. Two examples are considered: gauge and Yukawa theories. 

In both cases it is not sufficient to look at only a single graph; the UV divergence is eliminated only by summing up a proper set of graphs, or using the improved vertices which depend on our choice of bases for $S$. The choice of graphs depend on the graphs used to calculate the fermion self-energies.

\begin{figure*}[!t]
\begin{center}
\vspace{-1.5cm}
\hspace{-.3cm}
\includegraphics[width = 0.7\textwidth]{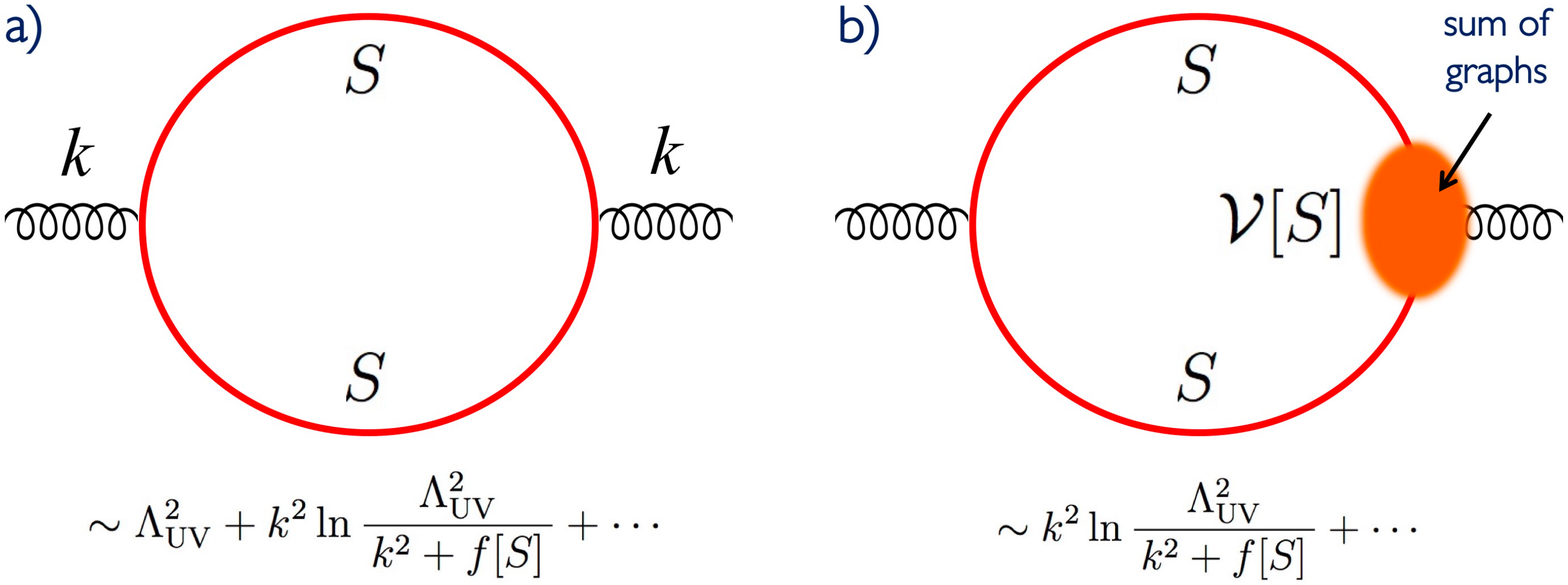}
\vspace{-1.5cm}
\end{center}
\vspace{-1.cm}
\caption{ 
\footnotesize{A gauge theory example for the fermion loop contributions to the boson self-energies. The fermion bases are $S (\neq S_{\rm tree} )$. The 1-loop result with a unimproved vertex results in the violation of the gauge invariance and quadratic divergence. They are associated with the change of bases $S_{\rm tree} \rightarrow S$.  Improving the vertex to recover the gauge invariance eliminates the quadratic divergence. 
}
}
\vspace{-0.0cm} 
\label{fig:vertex}
\end{figure*}

For gauge theories, the calculations are technically involved, but the required principle is clear-cut: the gauge invariance. The situation is schematically shown in Fig.\ref{fig:vertex}. The gauge invariance protects the polarization function from the quadratic divergences, and for this we need to include a proper set of 2PI-graphs which keeps the (truncated) 2PI functional gauge invariant. In particular the vertex must be improved as $V_{{\rm tree}} \rightarrow \calV[S]$, when we take into account the change of fermion bases \cite{Kojo:2014vja,Suenaga:2019jjv}. Once this is done for all $S$, the leading divergence of $\Pi^{[S]}$ always starts with the logarithmic one, and after single subtraction $\left( \Pi^{ [S ] } - \Pi^{ [S' ] }\right) $ becomes UV finite.

For Yukawa type theories, it is not easy to express the recipe in simple terms. Typically we have no symmetry restriction on the appearance of bosonic mass terms nor the quadratic divergence. After taking the difference between two self-energies, the quadratic divergence cancels, while the logarithmic divergence in general couples to quantities that depend on the fermion bases. Such logarithmic divergences must be combined with those appearing in the 4-point boson functions with fermion loops. (Unlike the gauge theory cases, the 2-point boson functions start with quadratic divergence so that 2-more boson insertions can reduce the divergence only to the logarithmic one.) In short, the renormalization program of 2-point boson functions requires resummed 4-point boson functions in the subgraphs. In the end the problem is reduced to the renormalization of the Bethe-Salpeter amplitudes appearing in the subgraphs \cite{vanHees:2001ik}. After this is done for all $S$, the single subtraction makes the boson self-energies UV finite.

In what follows, we will assume that we prepare the 2PI functional which includes the proper set of graphs to make the boson self-energies UV finite. 

Now we discuss the power counting of the difference of boson energies, $\Pi^{ [S ] } - \Pi^{ [S' ] }$, for different bases $S$ and $S'$. 
Before the subtraction the coefficients of $l^2$ contain the logarithmic divergence together with $l$-dependent functions dependent of the fermion bases,
\beq
\Pi^{ [S ] } (l) \sim c_{ {\rm univ} } \, l^2 \ln \frac{\, \Lambda_{{\rm UV}}^2 \,}{\, l^2 + f[S] \,} + O(l^0) \,.
\eeq 
where $f[S]$ is basis-dependent quantity of $\sim l^0$, and  $c_{ {\rm univ} } $ is the universal coefficient of the logarithmic divergence. 
At $l$ much larger than the nonperturbative scale ($l \gg f[S]$), subtracting the self-energy for $S'$ yields
\beq
\Pi^{ [S ] } (l) - \Pi^{ [S' ] } (l) \sim c_{ {\rm univ} } \, \left( f[S] - f[S'] \right)  \sim O(l^0) \,.
\eeq
As we have seen, this is necessary for the UV finiteness of the fermion self-energies. In next section we will also see that the same is required for the zero-point energy.

\subsection{The zero-point energy}

Finally we discuss the zero-point energy assembling the discussions given in the previous sections.

\begin{figure*}[!t]
\begin{center}
\vspace{-2.5cm}
\hspace{-.3cm}
\includegraphics[width = 0.8\textwidth]{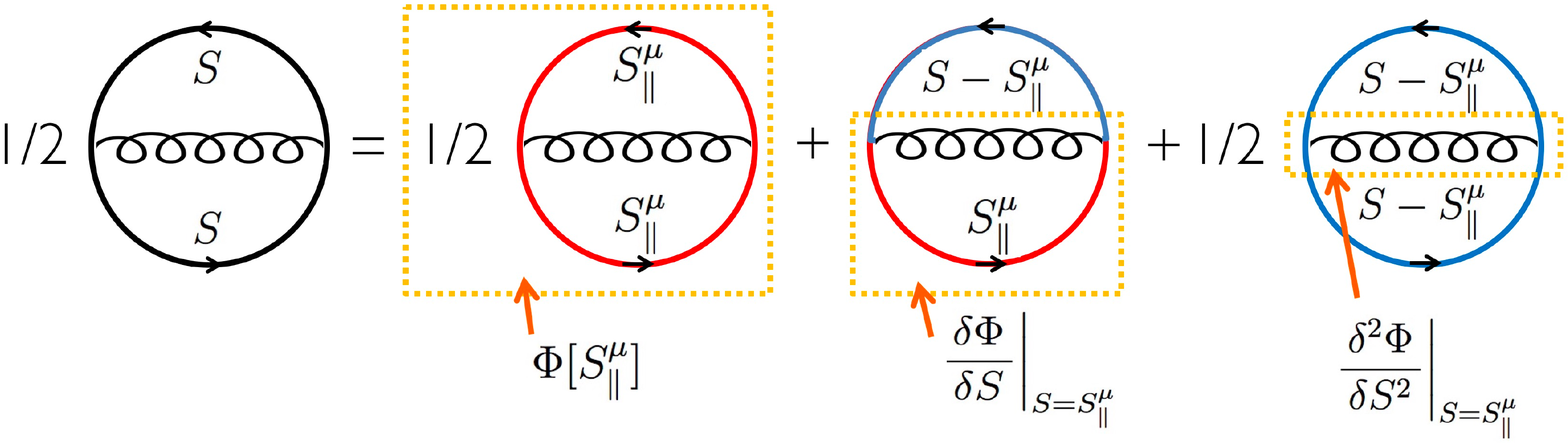}
\vspace{-0.cm}
\end{center}
\vspace{-3.5cm}
\caption{ 
\footnotesize{An example for the expansion of $\Phi [S]$ around $S=S_\para^\mu$. Shown is the Fock term.
}
\vspace{-0.0cm} }
\label{fig:expansion_phi}
\end{figure*}

Let us begin our analyses with the single particle contribution $\Tr \Ln S$. Its expansion starts from the linear order in $\Delta \Sigma$,
\beq
\Tr_1 \Ln S - \Tr_1 \Ln S^\mu_\parallel  = - \Tr_1 \Ln \left(1 + S^\mu_\parallel \Delta \Sigma \right) 
=  \sum_{n=1}  \frac{\, (-1)^n \,}{\, n \,} \Tr_1 \left[  \left( S^\mu_\parallel \Delta \Sigma \right)^n  \right] \,.
\label{eq:single}
\eeq
In our counting, the $n=1$ term is the order of $\sim \Lambda_{ {\rm UV} }^{2 + \gamma} $ that corresponds to the terms producing the quadratic divergences in Sec.\ref{sec:single}. The $n=2$ term is the order of $\sim  \Lambda_{ {\rm UV} }^{ 2\gamma}$ ($\sim \ln \Lambda_{ {\rm UV} }$ for $\gamma=0$), and the $n = 3$ terms are $\sim \Lambda_{ {\rm UV} }^{-2 + 3\gamma} $, and so on. (Clearly this expansion is useless for $\gamma \ge 2$.) Meanwhile the $\Tr_1 [S \tilde{\Sigma} ]$ terms in the 2PI action yield
\beq
\Tr_1 \left[ S \tilde{\Sigma}^{[S]} \right] - \Tr_1 \left[ S^\mu_\parallel \tilde{\Sigma}^{[S_\para]}  \right]
= \Tr_1 \left[ S^\mu_\parallel  \Delta \Sigma \right] + \Tr_1 \left[ \left( S -S^\mu_\parallel  \right) \tilde{\Sigma}^{[S]} \right] \,,
\eeq
$\tilde{\Sigma}^{ [S;\mu] } = S^{-1}- ( S^{\mu}_{ {\rm tree} } )^{-1}$
where we used $\tilde{\Sigma}^{[S_\para]} = S_\para^{-1}- ( S^{\mu}_{ {\rm tree} } )^{-1} = \Sigma_\para$. In our counting $S-S^\mu_\para \simeq - S_\para^\mu \left( \Delta \Sigma \right) S_\para^\mu \sim p^{-3+\gamma}$.
Here we observe that the first term in the RHS can be used to cancel the $n=1$ term in the $\Tr_1 \Ln S$ terms in Eq.(\ref{eq:single}), i.e., the strongest UV term in the single particle contribution. On the other hand the second term in the RHS is the order of $\Lambda_{ {\rm UV} }^{2 + \gamma}$. This piece can be cancelled by terms from the $\Phi$-functional, 
\beq
\Phi [S ] - \Phi [S^\mu_\para] 
= - \Tr_1\left[ \Sigma^{ [S^\mu_\para] } \left( S-S_\para^\mu \right) \right] 
+ \sum_{n=2} \frac{1}{\, n! \,} \Tr_n \left[ \frac{\, \delta^n \Phi \,}{\, \delta S^n \,} \bigg|_{S=S_\para^\mu} \left( S-S_\para^\mu \right)^n \right] \,,
\eeq
where we expand $\Phi[S]$ around $S^\mu_\para$. The meaning of the second trace is that we replace the $n$-propagators in the $\Phi$-functional by $S-S_\para^\mu$, and then set the rest of propagators to $S^\mu_\para$. As shown in Fig.\ref{fig:expansion_phi}, this is purely an algebraic operation: we can write $S = (S-S_\para^\mu) + S_\para^\mu$ and draw all possible diagrams made of $S-S_\para^\mu$ and $S_\para^\mu$. 

We note that the vertex functions $\delta^n \Phi /\delta S^n  \big|_{S=S_\para^\mu}$ are UV finite. Its evaluation can be done as performed in Sec.\ref{sec:self_same_bases}; we can isolate $\mu$-independent pieces of $n$-point vertex functions through the analytic continuation as in Eq.(\ref{formal_F2}),
\beq
\frac{\, \delta^n \Phi \,}{\, \delta S^n \,} \bigg|_{S=S_\para^\mu}
= \frac{\, \delta^n \Phi \,}{\, \delta S^n \,} \bigg|_{S=S_\para^{\mu=0} = S_*^{\mu=0} } + f_{ {\rm finite} }[\mu; S_*^{\mu=0} ]\,,
\eeq
where $f_{ {\rm finite} }[\mu; S_*^{\mu=0} ]$ is a functional of $\mu$ and $S_*^{\mu=0}$. Here are two remarks concerning with $\delta^n \Phi / \delta S^n $. First, its UV finiteness follows from the fact that $\delta^n \Phi / \delta S^n $ at $S_*^{\mu=0}$ is the vacuum $n$-point vertices which are UV finite. Second, taking derivative with respect to $S$, at each time the integration over loop momentum disappears while the $S$ in the denominator adds one power of momentum, so the powers of $\delta^n \Phi / \delta S^n $ are $\sim p^{4-3n}$. 

Assembling the above-mentioned three pieces of contributions to $I_{\Delta S}$, we find
\begin{align}
I_{\Delta S} [S;\mu] 
& = \sum_{n=2} \frac{\, (-1)^{n} \,}{\, n \,} \Tr_1 \left[  \left( S^\mu_\parallel \Delta \Sigma \right)^n  \right] 
 + \Tr_1 \left[ \left( S -S^\mu_\parallel  \right) \left( \tilde{\Sigma}^{[S;\mu]} -  \Sigma^{ [S^\mu_\para] } \right)  \right] 
 \nonumber \\
 & ~~~~ 
 +   \sum_{n=2}  \frac{1}{\, n! \,} \Tr_n \left[ \frac{\, \delta^n \Phi \,}{\, \delta S^n \,} \bigg|_{S=S_\para^\mu} \left( S-S_\para^\mu \right)^n \right] \,.
 \label{eq:result_2PI}
 \end{align}
The replacement of $S$ with $S-S_\para^\mu$ accompanies the reduction of the powers from $p^{-1}$ to $ \sim p^{-3+\gamma}$, so the terms in the last sum of the RHS is $\sim \Lambda_{ {\rm UV} }^{4-n(2 -\gamma) }$ and for $\gamma< 2$ the divergence is at most $\sim \Lambda_{ {\rm UV} }^{ 2\gamma }$.

We note that the worst divergences, $\sim \Lambda_{ {\rm UV} }^{2 + \gamma} $, have been cancelled. Indeed the powers of $\Delta \Sigma$ start from the quadratic order, $\sim (\Delta \Sigma)^2$, in which the powers are reduced by $p^{4-2\gamma}$. In the context of the quadratic divergences discussed in Sec.\ref{sec:pair}  and \ref{sec:single} (which were discussed with the assumption of $\gamma=0$), we first add them, and then subtract the double counted contributions; this results in the cancellation of the worst divergences. For this cancellation it is essential to keep track all the effects associated with the change of fermion bases; some appears from the single particle contributions and the other from composite particles or backgrounds. Now the leading divergence is weaker than naive expectations, and is the order of $\sim \Lambda_{ {\rm UV} }^{ 2\gamma}$ which becomes $\sim \ln \Lambda_{ {\rm UV} }$ at $\gamma = 0$. 

We can proceed further. From now on we focus on our canonical case, $\gamma \simeq 0$, and examine the logarithmic divergence. We pick up the UV divergent pieces in Eq.(\ref{eq:result_2PI})
 as
\begin{align}
I_{\Delta S} [S;\mu] 
& \simeq - \frac{\, 1 \,}{\, 2 \,} \Tr_1 \left[  \left( S  -S^\mu_\parallel  \right)  \left( \tilde{\Sigma}^{[S;\mu]} -  \Sigma_\para^\mu  \right)   \right] 
 + \Tr_1 \left[ \left( S -S^\mu_\parallel  \right) \left( \tilde{\Sigma}^{[S;\mu]} -  \Sigma^{ [S^\mu_\para] } \right)  \right] 
 \nonumber \\
 & ~~~~ 
 - \frac{1}{\, 2 \,} \Tr_2 \left[ \frac{\, \delta \Sigma \,}{\, \delta S \,} \bigg|_{S=S_\para^\mu} \left( S-S_\para^\mu \right)^2 \right] \,.
 \label{eq:result_2PI}
 \end{align}
where we used  $ S_\para^\mu \left( \Delta \Sigma \right) S_\para^\mu \simeq - (S-S^\mu_\para) $, $\Delta \Sigma^{ [S] } =  \tilde{ \Sigma}^{ [S] }  - \Sigma^\mu_\para $, and $\delta^2 \Phi/\delta S^2 = - \delta \Sigma/\delta S$.

So far we have not used specific properties of $S$ except its asymptotic behaviors. Now we substitute the solution of the Schwinger-Dyson equation, $S_*^\mu$, in place of $S$. Then, by definition $ \tilde{ \Sigma}^{ [S_*^\mu ] } = ( S_*^\mu)^{-1}- ( S^{\mu}_{ {\rm tree} } )^{-1} = \Sigma^{ [S_*^\mu ] }$, and we obtain
\begin{align}
I_{\Delta S} [S_*^\mu ;\mu] 
& \simeq - \frac{\, 1 \,}{\, 2 \,} \Tr_1 \left[  \left( S_*^\mu  -S^\mu_\parallel  \right)  \left( \Sigma^{[S_*^\mu]} -  \Sigma_\para^\mu  \right)   \right] 
 + \Tr_1 \left[ \left( S_*^\mu  -S^\mu_\parallel  \right) \left( \Sigma^{ [S_*^\mu ]} -  \Sigma^{ [S^\mu_\para] } \right)  \right] 
 \nonumber \\
 & ~~~~ 
 - \frac{1}{\, 2 \,} \Tr_2 \left[ \frac{\, \delta \Sigma \,}{\, \delta S \,} \bigg|_{S=S_\para^\mu} \left( S_*^\mu -S_\para^\mu \right)^n \right] \,.
 \label{eq:result_2PI}
 \end{align}
Finally we expand 
\beq
\Sigma^{[S_*^\mu]}  = \Sigma^{[S_\para^\mu]} + \frac{\, \delta \Sigma \,}{\, \delta S \,} \bigg|_{S=S_\para^\mu}  \left( S_*^\mu -S_\para^\mu \right) + \cdots
\eeq
with which we end up with
\beq
I_{\Delta S} [S_*^\mu ;\mu]  = - \frac{\, 1 \,}{\, 2 \,} \Tr_1 \left[  \left( S_*^\mu  -S^\mu_\parallel  \right)  \left( \Sigma^{[S_\para^\mu]} -  \Sigma_\para^\mu  \right)   \right] + O(S_*-S_\para)^3 \,.
\label{eq:result_final}
\eeq
The last term yields $\sim \Lambda_{ {\rm UV} }^{-2+3\gamma}$ and will be neglected. We could not find a way to further simplify the expression of Eq.(\ref{eq:result_final}), so at this point we stop transforming it, and then look into more details.  

\begin{figure*}[!t]
\begin{center}
\vspace{-2.5cm}
\hspace{-.3cm}
\includegraphics[width = 0.8\textwidth]{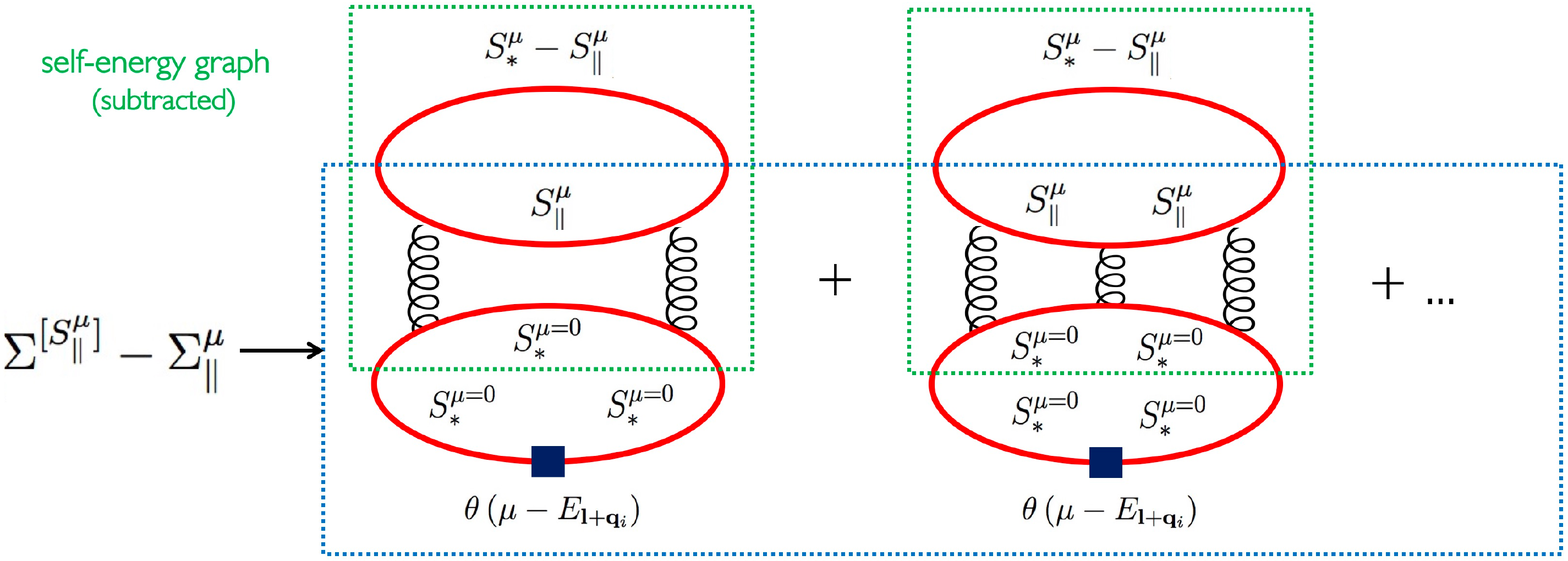}
\vspace{-0.cm}
\end{center}
\vspace{-2.5cm}
\caption{ 
\footnotesize{A 1-fermion loop example of Eq.(\ref{eq:result_final}). The self-energy difference $ \Sigma^{[S_\para^\mu]} -  \Sigma_\para^\mu $ is indicated by a box with blue lines. It is contracted with the subtracted propagator $S_*^\mu  -S^\mu_\parallel $. The box with green lines indicates the self-energy graph with one subtraction. They have the same UV structure as graphs shown in Figs.\ref{fig:2loop_s-s_type2} and \ref{fig:3loop_examplei} and are UV finite. As a result these graphs are UV finite after contracting with the $\theta$-function.
}
\vspace{-0.0cm} }
\label{fig:last_graph}
\end{figure*}

First we recall that the structure of $ \Sigma^{[S_\para^\mu]} -  \Sigma_\para^\mu $ is given by Eq.(\ref{eq:self-energy_general}) or Fig.\ref{fig:dif_self_energy}. Therefore Eq.(\ref{eq:result_final}) yields graphs shown in Fig.\ref{fig:last_graph} where we pick up graphs with 1-fermion loop as examples. If we decompose the graph into the $\theta$-function part and the others, the latter can be regarded as the self-energy graph in which single subtraction is done for the fermion loop. We have analyzed the UV structure of these graphs in Sec.\ref{sec:fermion-self} looking at graphs in Figs.\ref{fig:2loop_s-s_type2} and \ref{fig:3loop_examplei}. We concluded these graphs to be finite (after the vertex improvement). Contracting these UV finite self-energies with the $\theta$-function leaves UV finite results.

With this we conclude that the $I_{\Delta S}$ is UV finite, as far as the propagator $S$ leads to the UV finite self-energies for fermions and bosons.

\section{Summary}
\label{sec:summary}

We have investigated the UV divergences in the zero-point energy of composite particles that are associated with the change of fermion bases. We use the 2PI functional to keep track all the impacts due to the change of bases, and show that the UV divergences all cancel, provided that the in-medium fermion propagator at high energy approaches sufficiently fast to the vacuum counterpart.

While we started with the examples of 2-particle correlation or composite particles made of 2-particles, in the discussion of 2PI functional we did not refer to the specific form of the $\Phi$-functional, so the discussion can accommodate also the 3-, 4-, and infinite particle correlations.

In phenomenology, the present discussions may be important for the QCD equation of state at high baryon density where baryons merge exhibiting quark degrees of freedom. The relevant effective degrees of freedom change, and there must be a framework which simultaneously treat baryons and quarks while avoid the double counting. The 2PI formalism allows us such computation and cancels the apparent UV divergences which are expected from naive quasi-particle pictures for the constituents and composites.

We emphasize again that our attention in this paper is not necessarily restricted to renormalizable theories. 
Rather our intention is to understand the UV structure in general context so that we will be able to construct models or invent efficient cutoff schemes for effective theory calculations with physically motivated UV cutoff. After the removal of the artificial UV divergences, only the dependence on the physical cutoff remains so that our task is reduced to quantifying such cutoff scale from the microscopic calculations.
The practical implementation of composite particles will be presented elsewhere.

\section*{Acknowledgments}
This work was supported by NSFC grant 11650110435 and 11875144.
\appendix



\begin{thebibliography}{00}

\bibitem{bohr}
B. R. Mottelson and A. N. Bohr, {\it Nuclear Structure}, World Scientific, 1998.

\bibitem{Nambu:1961tp}
  Y.~Nambu and G.~Jona-Lasinio,
  Phys.\ Rev.\  {\bf 122} (1961) 345;
 J. Goldstone, A. Salam, and S. Weinberg, Phys. Rev. {\bf 127} (1962), 965.
 
 \bibitem{Higgs} 
Y.~Nambu, 
 Phys.\ Rev.\  {\bf 117} (1960) 648;
 P. W. Anderson, Phys. Rev. {\bf 130} (1962) 439.
  
\bibitem{AGD}
A. A. Abrikosov, L. P. Gor'kov and I. E. Dzyaloshinski, {\it Methods of Quantum Field Theory in Statistical Physics}, Dover Books on Physics.

\bibitem{composite_fermon}
J. K. Jain, {\it Composite Fermions}, Cambridge University Press, 2007.


\bibitem{Schafer:1998ef}
  T.~Sch$\ddot{ {\rm a} }$fer and F.~Wilczek,
  Phys.\ Rev.\ Lett.\  {\bf 82} (1999) 3956.

\bibitem{Baym:2019iky}
  G.~Baym, S.~Furusawa, T.~Hatsuda, T.~Kojo and H.~Togashi,
  Astrophys. J. {\bf 885} (2019), no. 1.
\bibitem{Baym:2017whm}
For recent reviews on quark matter, e.g.,
 G.~Baym, T.~Hatsuda, T.~Kojo, P.~D.~Powell, Y.~Song and T.~Takatsuka,
  Rept.\ Prog.\ Phys.\  {\bf 81} (2018) no.5,  056902;
  T.~Kojo,
  Eur.\ Phys.\ J.\ A {\bf 52} (2016) no.3, 51.

\bibitem{Masuda:2012kf}
  K.~Masuda, T.~Hatsuda and T.~Takatsuka,
  Astrophys.\ J.\  {\bf 764} (2013) 12.

\bibitem{Kojo:2014rca}
  T.~Kojo, P.~D.~Powell, Y.~Song and G.~Baym,
  Phys.\ Rev.\ D {\bf 91} (2015) no.4,  045003.

\bibitem{Fukushima:2015bda}
  K.~Fukushima and T.~Kojo,
  Astrophys.\ J.\  {\bf 817} (2016) no.2,  180.
\bibitem{McLerran:2018hbz}
  L.~McLerran and S.~Reddy,
  Phys.\ Rev.\ Lett.\  {\bf 122} (2019) no.12,  122701.
\bibitem{Jeong:2019lhv}
  K.~S.~Jeong, L.~McLerran and S.~Sen,
  arXiv:1908.04799 [nucl-th].

\bibitem{Ma:2019ery}
  Y.~L.~Ma and M.~Rho,
  arXiv:1909.05889 [nucl-th].

  
  

\bibitem{Dashen:1969ep}
  R.~Dashen, S.~K.~Ma and H.~J.~Bernstein,
  Phys.\ Rev.\  {\bf 187} (1969) 345.


\bibitem{Blaschke:2013zaa}
  D.~Blaschke, M.~Buballa, A.~Dubinin, G.~R$\ddot{ \rm o}$pke and D.~Zablocki,
  Annals Phys.\  {\bf 348} (2014) 228.

\bibitem{Bastian:2018wfl}
  N.~U.~F.~Bastian, D.~Blaschke, T.~Fischer and G.~R$\ddot{ \rm o}$pke,
  Universe {\bf 4} (2018) 67.

\bibitem{Kojo:2017gxc}
  T.~Kojo,
  PoS CPOD {\bf 2017} (2018) 071.
\bibitem{Kojo:2017opq}
  T.~Kojo,
  Nucl.\ Phys.\ A {\bf 967} (2017) 832.

\bibitem{Freedman:1976xs}
  B.~A.~Freedman and L.~D.~McLerran,
  Phys.\ Rev.\ D {\bf 16} (1977) 1130;
  {\it ibid.} 1147;
  {\it ibid.} 1169.

\bibitem{Kurkela:2009gj}
  A.~Kurkela, P.~Romatschke and A.~Vuorinen,
  Phys.\ Rev.\ D {\bf 81} (2010) 105021.

\bibitem{Xia:2019xax}
  C.~Xia, Z.~Zhu, X.~Zhou and A.~Li,
  arXiv:1906.00826 [nucl-th].

\bibitem{Han:2019bub}
  S.~Han, M.~A.~A.~Mamun, S.~Lalit, C.~Constantinou and M.~Prakash,
  arXiv:1906.04095 [astro-ph.HE].

\bibitem{NSR}
P. Nozieres and S. Schmitt-Rink, J. Low. Temp. Phys. {\bf 59} (1985) 195.

\bibitem{Diener}
R. B. Diener, R. Sensarma, and M. Randeria, 
Phys.\ Rev.\ A {\bf 77} (2008) 023626.
   
\bibitem{Ohashi:2002zz}
  Y.~Ohashi and A.~Griffin,
  Phys.\ Rev.\ Lett.\  {\bf 89} (2002) 130402.
 
\bibitem{Abuki:2006dv}
  H.~Abuki,
  Nucl.\ Phys.\ A {\bf 791} (2007) 117.
 
\bibitem{Sun:2007fc}
  G.~f.~Sun, L.~He and P.~Zhuang,
  Phys.\ Rev.\ D {\bf 75} (2007) 096004.
    
\bibitem{Hatsuda:1994pi}
  T.~Hatsuda and T.~Kunihiro,
  Phys.\ Rept.\  {\bf 247} (1994) 221.
    
\bibitem{Weinberg}
S. Weinberg, {\it The Quantum Theory of Fields}, Cambridge University Press, 2005, Chapter 12.    
\bibitem{Peskin}
M. E. Peskin and D. V. Schroeder,
{\it An Introduction to quantum field theory},  Addison-Wesley, 1995.

\bibitem{Kapusta}
J. I. Kapusta and C. Gale, {\it Finite-Temperature Field Theory: Principles and Applications}, Cambridge University Press, 2006.


\bibitem{Luttinger:1960ua}
  J.~M.~Luttinger and J.~C.~Ward,
  Phys.\ Rev.\  {\bf 118} (1960) 1417.


\bibitem{Baym:1961zz}
  G.~Baym and L.~P.~Kadanoff,
  Phys.\ Rev.\  {\bf 124} (1961) 287.

\bibitem{Baym:1962sx}
  G.~Baym,
  Phys.\ Rev.\  {\bf 127} (1962) 1391.
    
\bibitem{Cornwall:1974vz}
  J.~M.~Cornwall, R.~Jackiw and E.~Tomboulis,
  Phys.\ Rev.\ D {\bf 10} (1974) 2428.

\bibitem{vanHees:2001ik}
  H.~van Hees and J.~Knoll,
  Phys.\ Rev.\ D {\bf 65} (2002) 025010;
{\it ibid.} 
  105005;
  {\bf 66} (2002) 025028.

\bibitem{Blaizot:2003an}
  J.~P.~Blaizot, E.~Iancu and U.~Reinosa,
  Nucl.\ Phys.\ A {\bf 736} (2004) 149;
{\it ibid.} 
  Phys.\ Lett.\ B {\bf 568} (2003) 160;
  U.~Reinosa and J.~Serreau,
  Annals Phys.\  {\bf 325} (2010) 969.

\bibitem{Weinberg:1978kz}
  S.~Weinberg,
  Physica A {\bf 96} (1979) no.1-2,  327.


\bibitem{Coleman:1973jx}
  S.~R.~Coleman and E.~J.~Weinberg,
  Phys.\ Rev.\ D {\bf 7} (1973) 1888.

\bibitem{Kojo:2014vja}
  T.~Kojo and G.~Baym,
  Phys.\ Rev.\ D {\bf 89} (2014) no.12,  125008.

\bibitem{Suenaga:2019jjv}
  D.~Suenaga and T.~Kojo,
  Phys.\ Rev.\ D {\bf 100} (2019) no.7,  076017.
  










    
\end{thebibliography}
\end{document}